\documentclass{article} % For LaTeX2e
\usepackage{iclr2022_conference,times}

\usepackage{amsmath,amsfonts,bm}

\def\eqref#1{equation~\ref{#1}}
\def\1{\bm{1}}

\DeclareMathAlphabet{\mathsfit}{\encodingdefault}{\sfdefault}{m}{sl}
\SetMathAlphabet{\mathsfit}{bold}{\encodingdefault}{\sfdefault}{bx}{n}

\usepackage{wrapfig}
\usepackage{hyperref}
\usepackage{url}
\usepackage{graphicx}
\usepackage{tabularx}
\usepackage{multirow}
\usepackage{graphicx}
\usepackage{caption}
\usepackage{newunicodechar}

\usepackage{subcaption}
\usepackage{stfloats} % for positioning of figure* on the same page
\usepackage{lipsum}
\title{AnoSeg: Anomaly Segmentation Network Using Self-Supervised Learning}

\author{Jou Won Song{$^1$}\thanks{*equal contribution} , Kyeongbo Kong{$^{2\star}$}, Ye-In Park{$^1$}, Seong-Gyun Kim{$^3$}, Suk-Ju Kang{$^1$} \\
{$^1$}Department of Electronic Engineering, Sogang University, Seoul, Korea\\
{$^2$}Department of Media communication, Pukyong National University, Busan, Korea\\
{$^3$}LG Display, Seoul, South Korea\\
\texttt{\{wn5649,yipark,sjkang\}@sogang.ac.kr}{$^1$} \\
\texttt{\{kbkong\}@pknu.ac.kr}{$^2$} \\
\texttt{\{ksglcd\}@lgdisplay.com}{$^3$} \\

}

\iclrfinalcopy % Uncomment for camera-ready version, but NOT for submission.
\begin{document}

\maketitle

\begin{abstract}
Anomaly segmentation, which localizes defective areas, is an important component in large-scale industrial manufacturing. However, most recent researches have focused on anomaly detection. This paper proposes a novel anomaly segmentation network (AnoSeg) that can directly generate an accurate anomaly map using self-supervised learning. For highly accurate anomaly segmentation, the proposed AnoSeg considers three novel techniques: Anomaly data generation based on hard augmentation, self-supervised learning with pixel-wise and adversarial losses, and coordinate channel concatenation. First, to generate synthetic anomaly images and reference masks for normal data, the proposed method uses hard augmentation to change the normal sample distribution. Then, the proposed AnoSeg is trained in a self-supervised learning manner from the synthetic anomaly data and normal data. Finally, the coordinate channel, which represents the pixel location information, is concatenated to an input of AnoSeg to consider the positional relationship of each pixel in the image. The estimated anomaly map can also be utilized to improve the performance of anomaly detection. Our experiments show that the proposed method outperforms the state-of-the-art anomaly detection and anomaly segmentation methods for the MVTec AD dataset. In addition, we compared the proposed method with the existing methods through the intersection over union (IoU) metric commonly used in segmentation tasks and demonstrated the superiority of our method for anomaly segmentation.
\end{abstract}

\section{Introduction}

%As computer vision developed from image-level classification to pixel-level segmentation, anomaly detection methods have also developed into pixel-level anomaly segmentation after image-level anomaly detection was developed. 

Anomaly segmentation is the process that localizes anomaly regions. In the real world, since the number of anomaly data is very limited, conventional anomaly segmentation methods are trained using only normal data. Typically, many anomaly segmentation methods are based on anomaly detection techniques because the real dataset includes few anomaly images without the ground truth (GT) mask. Therefore, these methods are not trained directly on pixel-level segmentation and they are difficult to generate anomaly maps similar to GT masks.

Specifically, existing reconstruction-based methods using autoencoder (AE) (\cite{re8,re9,re12,re10, mvtec}) and generative adversarial network (GAN) (\cite{re7,re11,anog,re14}) are trained to learn reconstruction of normal images and determine anomaly if the test sample has the high reconstruction error for an abnormal region. However, reconstruction-based methods often restore even non-complex anomaly regions, which degrade the performance on both anomaly detection and segmentation. Therefore, the anomaly map in Fig. \ref{fig1}(b) greatly differs from the corresponding GT mask. Alternative methods have been recently studied by using the high-level learned representation for anomaly detection and segmentation. These methods use a pretrained model to extract a holistic representation of a given image and compare it to the representation of a normal image. Also, several existing methods use patches, splitting a given image to perform anomaly segmentation. By extracting representations from an image patch, these methods compute the scores of the image patches and combine them to generate the final anomaly map. Therefore, the quality of anomaly maps is highly correlated with the patch size. The uninformed students (US) (\cite{stu}) in Figs. \ref{fig1}(c) and (d) are trained using a small patch size (17 x 17) and a large patch size (65 x 65), respectively. Therefore, as shown in Fig. \ref{fig1}(d), US\textsubscript{65 x 65} is difficult to detect small anomaly regions. Patch SVDD (\cite{patch}) and SPADE (\cite{spa}) use feature maps of multiple scales to detect anomaly regions with various sizes. However, as shown in Figs. \ref{fig1}(e) and (f), these methods approximately detect anomaly regions. In addition, in GradCAM-based methods, GradCAM (\cite{grad}) is used to generate anomaly maps to detect regions that influence the decision of the trained model (\cite {att,eatt}). CutPaste (\cite{cut}) introduces a self-supervised framework using a simple effective augmentation that encourages the model to find local irregularities. CutPaste also performs anomaly localization through GradCAM by extending the model to use patch images after training the classifier. However, these methods are not aimed at anomaly segmentation and detect anomaly regions using a modified anomaly detection method. Generally, to improve the segmentation performance, a methodology that can be learned pixel-wise should be considered. Therefore, existing methods cannot clearly detect anomalies because it is difficult that directly use the pixel-wise loss such as a mean squared error typically used in the segmentation task.

To handle this problem, this paper proposes a new methodology that can directly learn the segmentation task. The proposed anomaly segmentation network (AnoSeg) can generate an anomaly map to segment the anomaly region that is unrelated to the normal class. The goal of AnoSeg is to generate an anomaly map that represents the normal class region within a given image for anomaly segmentation, unlike the existing methods to indirectly extract anomaly maps. For this goal, our AnoSeg proposes three following approaches. First, as shown in Fig. 2, AnoSeg uses the segmentation loss directly using the synthesized data generated through hard augmentation, which generates data shifted away from the input data distribution. Second, AnoSeg learns to generate the anomaly map and reconstruct normal images. 

Also, an adversarial loss is applied by using a generated anomaly map and an input image. Unlike the existing GAN, the discriminator of AnoSeg determines whether the image is a normal class and whether the anomaly map is focused on the normal region. Since the anomaly map learns the normal sample distribution, AnoSeg has high generalization for unseen normal and anomaly regions even with a small number of normal samples. 

Third, we propose the coordinate channel concatenation using a coordinate vector based on coordconv (\cite{coord}). Anomaly regions in a particular category often depend on the location information of a given image. Therefore, the proposed coordinate vector helps to understand the positional relationship of normal and anomaly regions in the input image. As a result, Fig. \ref{fig1}(h) shows that the anomaly map of AnoSeg is very similar to GT even without thresholding. Moreover, we describe how to perform the anomaly detection using the generated anomaly map. By simply extending the model using an anomaly map to the existing GAN-based method (\cite{alocc}), we could achieve 96.4 area under ROC curve (AUROC) for image-level localization, which is a significant improvement over conventional state-of-the-art (SOTA) methods. As a result, the proposed method achieves SOTA performance on the MVTec Anomaly Detection (MVTec AD) dataset for anomaly detection and segmentation compared to conventional methods without using a pretrained model. The main contributions of this study are summarized as follows:

\begin{figure*}[t]
\begin{center}
\includegraphics[width=0.95\linewidth]{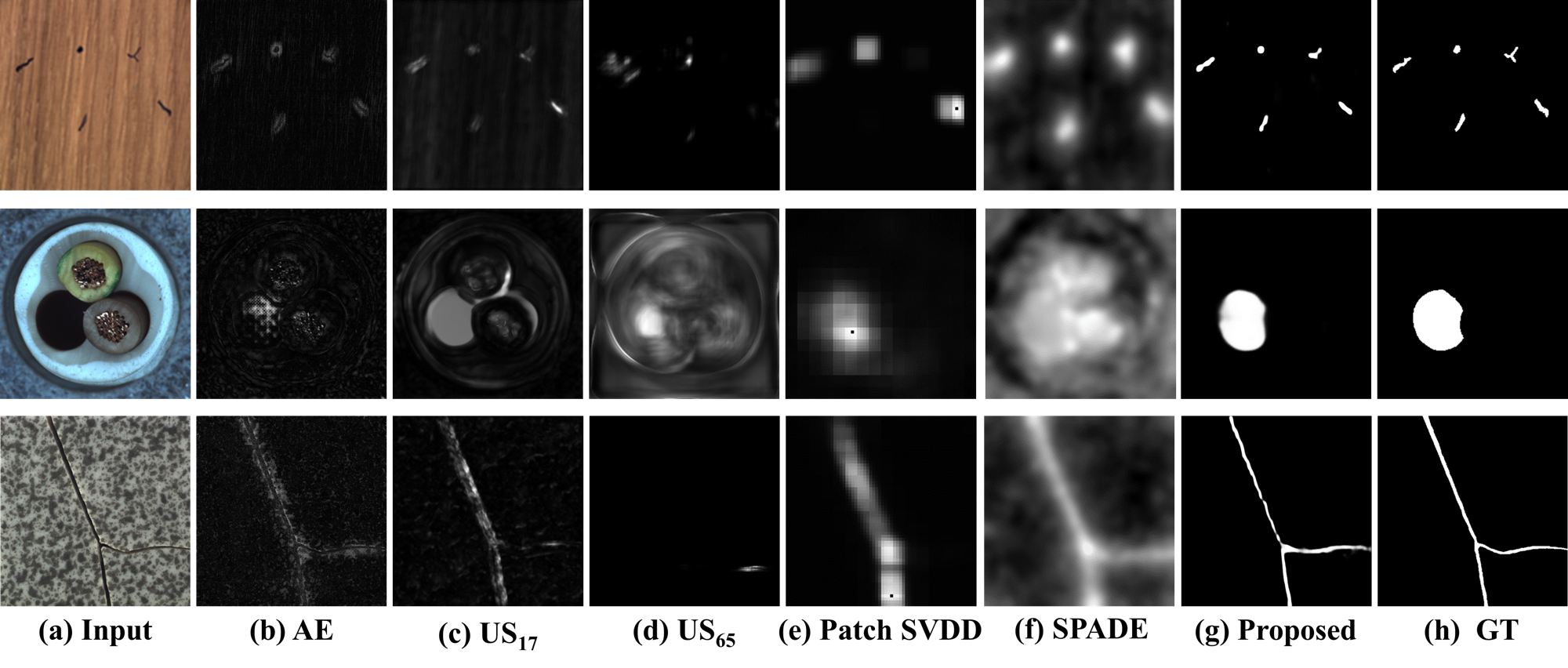} 
\end{center}
\vspace{-0.3cm}
   \caption{Comparison of anomaly maps (before thresholding) of the proposed method with the SOTA methods in the MVTec-AD dataset. Except for the proposed method, anomaly maps of existing methods are normalized to [0, 1].}
\label{fig1}
\vspace{-0.4cm}
\end{figure*}

\begin{itemize}
\item We propose a novel anomaly segmentation network (AnoSeg) to directly generate an anomaly map. AnoSeg generates detailed anomaly maps using the holistic approaches to maximize segmentation performance.
\item The proposed anomaly map can also be used in existing anomaly detection methods to improve the anomaly detection performance. 
\item In anomaly segmentation and detection, AnoSeg outperforms SOTA methods on the MVTec AD dataset in terms of intersection over union (IoU) and AUROC. Additional experiments using IoU metric also show that AnoSeg is robust for thresholding.
\end{itemize}

%As confirmed in [OE],[semi], using small amount of anomaly data or auxiliary out of distribution (OOD) datasets help to estimate the normal sample distribution. However, the auxiliary dataset must be similar to the normal class dataset, and it is difficult to obtain such datasets in a real case. Therefore, the proposed method uses hard augmented anomaly data extracted from normal data instead of real anomaly data to improve performance. As a result, the proposed method achieves state-of-the-art performance on mvtec ad dataset compared to conventional methods without using a pretrain model. Also, through evaluation on the CIFAR10 dataset, we demonstrate that the proposed method is effective for semantic anomaly detection.
% \begin{itemize}
% \item We propose a novel anomaly segmentation GAN to generate an anomaly map that focuses on the normal class region. Unlike existing methods, the proposed ASGAN generates detailed anomaly maps because it is trained to directly generate anomaly maps using hard augmentation.
% \item The two-stage network structure, which connects the ASGAN and ADGAN, is proposed  for sensitive anomaly detection. The ADGAN learns the normal class distribution of anomaly maps as well as input images, and hence, it can detect the anomaly class more sensitively than existing methods.
% \item The proposed method outperforms the state-of-the-art methods on several datasets in terms of the classification accuracy and the average area under a receiver operating characteristic curve (AUROC).
% \end{itemize}

\section{Related Works}
Anomaly detection is a research topic that has received considerable attention. Anomaly detection and segmentation are usually performed via unsupervised methods using the generative model for learning the distribution of a certain class. In these methods, GAN (\cite{gan}) or VAE (\cite{vae}) learned the distribution of a certain class and used the difference between a reconstructed image and an input for anomaly detection (\cite{re8,re10, re12,alocc}). In addition, initial deep learning-based anomaly segmentation methods focused on generative models such as GAN (\cite{anog}) and AE (\cite{mvtec}). However, these approaches could have high reconstruction performance for simple anomaly regions. Recently, methods using a representation of an image patch have shown great effectiveness in anomaly detection (\cite{patch, spa}). In \cite{stu}, US was trained to mimic a pretrained teacher by dividing an image into patches. In recent studies (\cite{cut}), an activation map that visualizes the region of interest through GradCAM (\cite{grad}) was applied to anomaly detection. \cite{att} generated an activation map using GradCAM to focus only on the reconstruction loss of the ROI. \cite{eatt} improved the detection performance using an activation map in the training process. \cite{fcdd} apply one-class classification on features extracted from a fully convolutional network and use receptive field upsampling with Gaussian smoothing to extract anomaly map. However, in these existing methods, it is difficult to apply the loss related to anomaly segmentation because the model does not directly generate an anomaly map by using the modified anomaly detection method. Our method is different from the conventional methods which use GradCAM to indirectly extract the activation map. Instead, the proposed method directly extracts and supervises the anomaly map. Therefore, the proposed method discriminates between anomaly and normal regions more accurately compared to previous methods.

\begin{figure}[t]
\begin{center}
\includegraphics[width=1.0\linewidth]{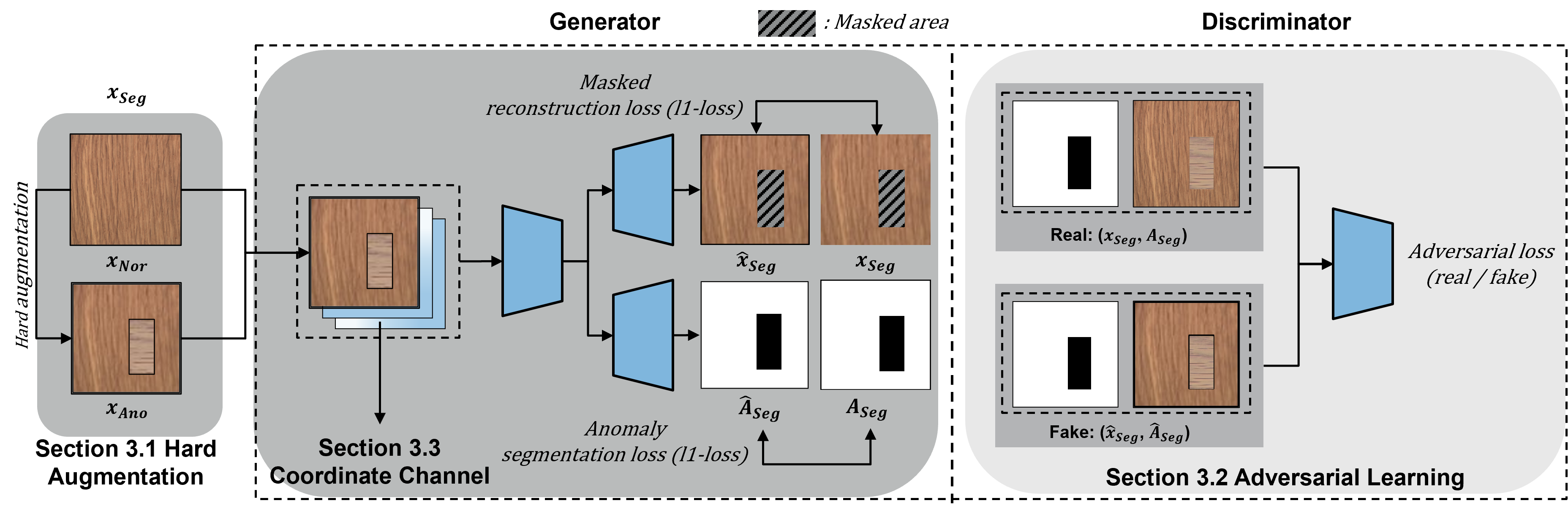} 
\end{center}
   \caption{Overview of the training process of the proposed AnoSeg. AnoSeg generates reconstructed images and anomaly maps. To directly generate anomaly maps, AnoSeg applies three novel techniques: hard augmentation, adversarial learning, and coordinate channel concatenation.}
\label{fig2}
\vspace{-0.4cm}
\end{figure}

% \subsection{Hard Augmentation}
% Hard augmentation refers to generating samples shifted away from the original sample distribution. The augmentation is not commonly used because the label of the image may not be preserved. In \cite{20,24,4}, geometric transformations, such as rotation, translation or flips, is predicted for anomaly detection. \cite {csi} proposed to push out samples with hard or distribution-shifting augmentations from the original sample and validate the hard augmentation effect for anomaly detection. Cutpaste (\cite{cut}) cut a small rectangular area of variable sizes from a normal training image and paste a patch back to an image at a random location to generate synthetic anomaly data. In this paper, a synthetic anomaly data is generated by performing Cutpaste on the data to which hard augmentation is applied. Therefore, AnoSeg is trained on a various anomaly regions, which improves anomaly segmentation performance.

\section{Proposed Method: AnoSeg}
The proposed AnoSeg is a ``holistic'' approach which incorporates three techniques: self-supervised learning using hard augmentation, adversarial learning, and coordinate channel concatenation. The details are explained in the following sub-sections.

%As shown in Fig. 2, the overall structure of AnoSeg consists of a generator that estimates anomaly map and reconstructed images, and a discriminator for adversarial learning. To directly generate anomaly maps using self-supervised learning, AnoSeg considers three techniques: Anomaly data generation using hard augmentation, adversarial loss for learning the normal sample distribution, and position vector concatenation. The details are explained in the following sections.

\subsection{Self-supervised Learning Using Hard Augmentation}
To train anomaly segmentation directly, an image with an anomaly region and its corresponding GT mask corresponding to the image are required. However, it is difficult to obtain these images and GT masks in the real case. Therefore, the proposed method uses hard augmentation (\cite{csi}) and Cutpaste (\cite{cut}) to generate synthetic anomaly data and GT masks. Hard augmentation refers to generating samples shifted away from the original sample distribution. As confirmed in \cite{csi}, the hard augmented samples can be used as a negative samples. Therefore, as shown in Fig. 3, we use three types of hard augmentation: rotation, perm, and color jitter. Each augmentation is applied with a 50\% chance. Then, like Cutpaste (\cite{cut}), the augmented data is pasted into a random region of normal data to generate the synthetic anomaly data and corresponding masks for segmentation. Finally, the anomaly segmentation dataset is composed as follows:
\begin{equation}
  x_{Seg}=\left\{x_{Nor}, x_{Ano}\right\}, A_{Seg}=\left\{A_{Nor}, A_{Ano}\right\},
  \label{equ:seg_data}
\end{equation}
where $x_{seg}$ is a set of normal and synthetic anomaly images, in which $x_{Nor}$ and $x_{Ano}$ are normal images and synthetic anomaly images, respectively. $A_{seg}$ is a set of normal and synthetic anomaly masks, in which $A_{Nor}$ and $A_{Ano}$ are normal masks with all inner values set to one and synthetic anomaly masks, respectively.

Using the anomaly segmentation dataset with a pixel-level loss, we can directly train our AnoSeg. The anomaly segmentation loss $L_{Seg}$ is as follows:
\begin{equation}
  L_{Seg} = \mathbb{E}\parallel A_{Seg}-\,\widehat{A}_{Seg} \parallel ^{1},
  \label{equ:dt}
\end{equation}
where $\widehat{A}_{Seg}$ indicates the generated anomaly map (normal and anomaly classes). The generated anomaly map has the same size as an input image and outputs a value in the range of [0, 1] for each pixel depending on the importance of the pixel of the input image. However, since the synthetic anomaly data are only subset of various anomaly data, it is difficult to generate a real anomaly maps that are unseen in training phase.

%Reconstructed images are used for adversarial learning in the next section.

\begin{figure}[t]
\begin{center}
\includegraphics[width=1.0\linewidth]{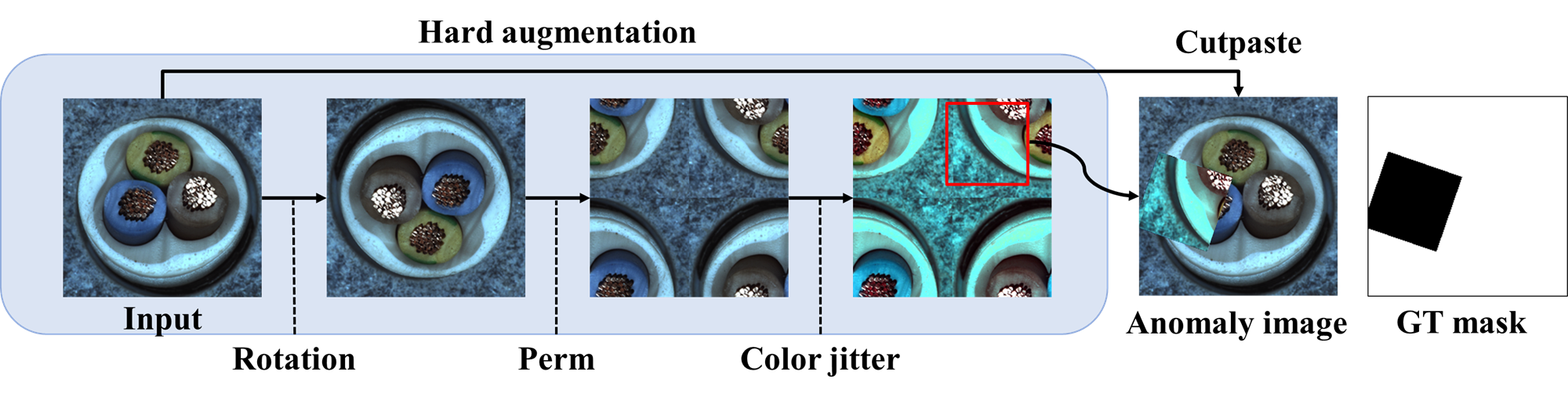} 

\end{center}
\vspace{-0.2cm}
\caption{Our synthetic anomaly data augmentation. The synthetic anomaly data is generated by several hard augmentations and Cutpaste (\cite{cut}). Synthetic anomaly data is generated by applying a rotation, perm, color jitter, and Cutpaste for each step. Hard augmentations are applied with a 50\% chance.}
\vspace{-0.2cm}
\label{fig3}
\end{figure}
\subsection{Adversarial Learning with Reconstruction}
To improve the generality for various anomaly data, it is important to train normal region distribution accurately. Therefore, AnoSeg utilizes masked reconstruction loss that uses reconstruction loss only in normal regions to learn only the distribution of normal regions and avoid bias of the distribution of synthetic anomaly regions. Also, since the discriminator inputs a pair for an input image and its GT masks, the discriminator and generator can focus on normal region distribution. Thus, anomaly region cannot be reconstructed well and the detail of the anomaly map can also be improved. Loss functions for adversarial learning are as follows:
\begin{align}
  L_{Adv} = \underset{G}{min} \underset{D}{max}\{\mathbb{E}\;[\log(D(concat(x_{Seg},A_{Seg})))]+\mathbb{E}\;[\log(1-D(concat(\widehat{x}_{Seg},\widehat{A}_{Seg})))]\},
  \label{equ:dt}
\end{align}
\begin{equation}
   L_{Re} = \mathbb{E}\parallel x_{Seg}*A_{Seg}-\,\widehat{x}_{Seg}*A_{Seg} \parallel ^{1}/\mathbb{E}\parallel A_{Seg}\parallel ^{1},
   \label{equ:dt}
\end{equation}
where $D$, $G$, and $concat$ are a discriminator, a generator, and a concatenation operation, respectively. In Section 5, we demonstrated the effectiveness of adversarial loss.
%Therefore, AnoSeg utilizes image reconstruction and adversarial learning which is one of the best ways to learn data distribution. The discriminator inputs a pair of input images and GT mask containing both normal and synthetic anomaly data. Thus, the discriminator learns the distribution of pairs, and generator of AnoSeg is trained to generate anomaly maps that segment the abnormal and normal regions. In addition, we use masked reconstruction losses that use reconstruction losses only in normal regions to learn only the distribution of normal regions and avoid bias in the distribution of synthetic anomaly regions. Loss functions for adversarial learning are as follows:

%Therefore, the proposed anomaly map detect anomaly regions in detail without being biased toward a specific anomaly region.

% In addition, to better represent the normal class distribution, the reconstruction loss of the normal image is used in the generator of AnoSeg. 

% The loss function for learning the generator are as follows:
% \begin{align}
%  L_{G} = L_{R}(x_{Nor}, x_{Nor}^{'}),
%   \label{equ:dt}
% \end{align}
% where $L_{R}$, $x_{Nor}^{'}$, and $x_{Nor}$ is mean squared error, a reconstructed normal image, and a normal image, respectively.

%\begin{wrapfigure}{r}{0.6\textwidth}
%\centerline{\includegraphics[width=8  cm]{44.png}}
% \caption{Overall process of the coordinate channel concatenation.}
%\label{fig:wrapfig}
%\end{wrapfigure}

\begin{wrapfigure}{H}{0.5\textwidth}
\hspace{-10pt}
 \begin{center}
 \vspace{-12pt}
 \centerline{\includegraphics[width=0.5\columnwidth]{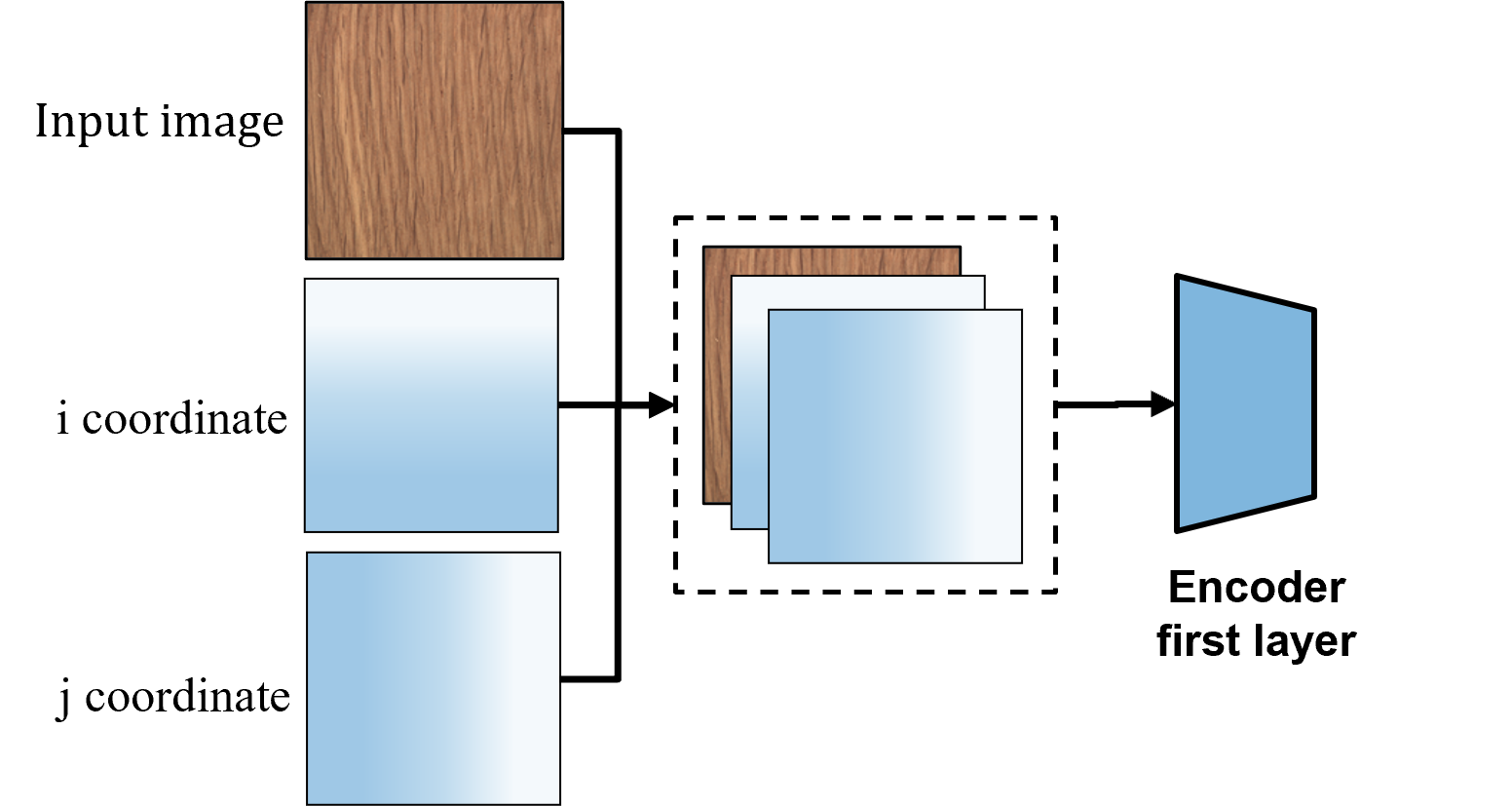}}
 \end{center}
 \vspace{-20pt}
 \caption{Overall process of the coordinate channel concatenation.}
\label{fig4}
 \vspace{-10pt}
\end{wrapfigure}

% \begin{figure}[b]
% \begin{center}
% \includegraphics[width=1.0\linewidth]{55.png} 
%   %\includegraphics[width=0.8\linewidth]{egfigure.eps}
% \end{center}
%   \caption{An overview of the proposed anomaly detection method. }
% \label{fig2}
% %\vspace{-0.4cm}
% \end{figure}
%In the first step, AnoSeg generates an anomaly map of the input image. In the next step, the pair of images reconstructed from the anomaly map and the anomaly detector are compared with the pair of the normal mask and the input image using a discriminator of detector.

\subsection{Coordinate Channel Concatenation}
In the typical segmentation task, the location information is the most important information because normal and anomaly regions can be changed depending on where they are located. To provide additional location information, we use a coordinate vector inspired by CoordConv (\cite{coord}). We first generate rank 1 matrices that are normalized to [-1, 1]. Then, we concatenate these matrices with the input image as channels (Fig. \ref{fig4}). As a result, AnoSeg extracts features by considering the positional relationship of the input image. In ablation study, we demonstrated the effectiveness of coordinate channel concatenation.

\begin{wrapfigure}{H}{0.5\textwidth}
\hspace{-10pt}
 \begin{center}
 \vspace{-20pt}
 \centerline{\includegraphics[width=0.5\columnwidth]{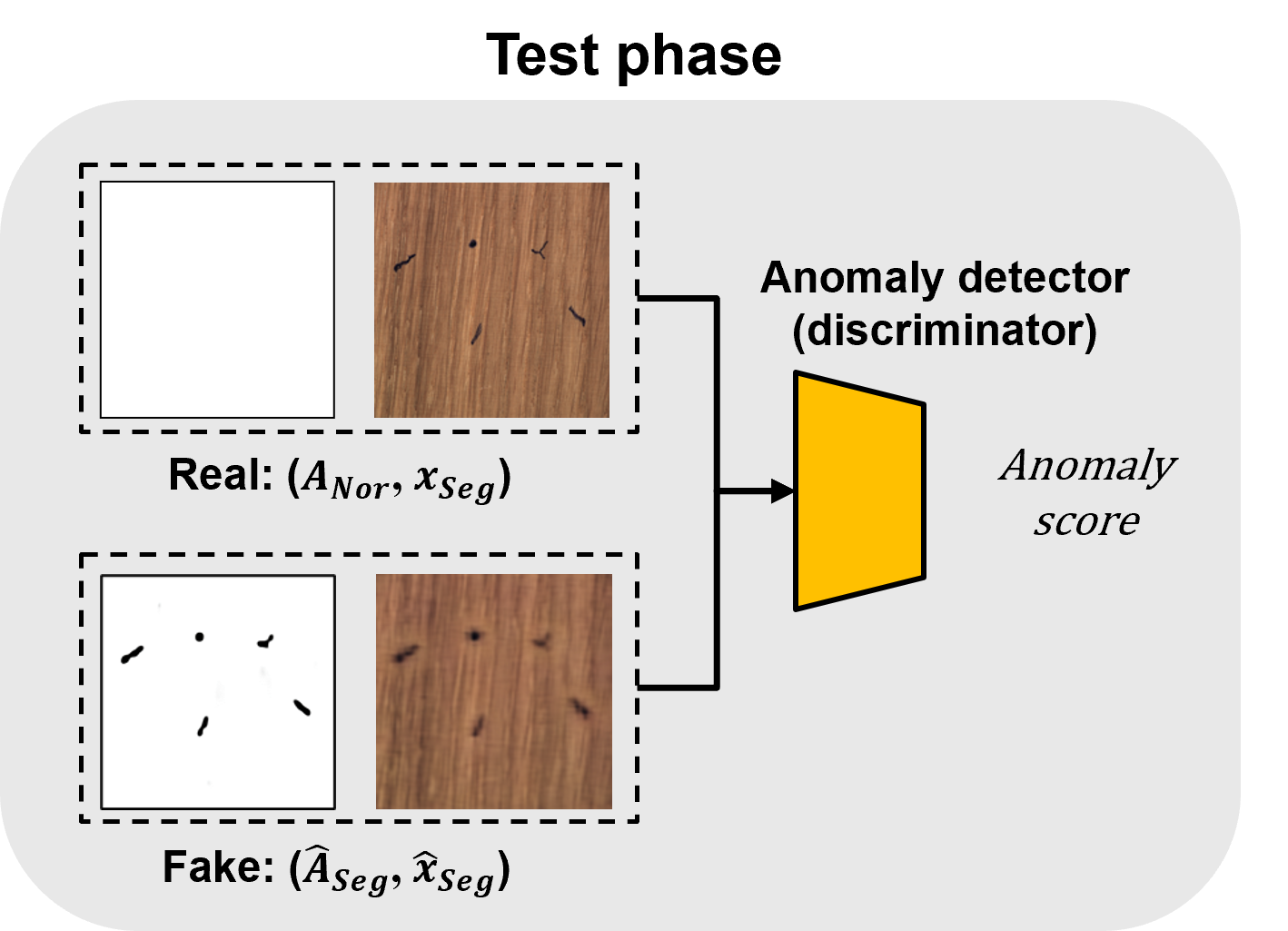}}
 \end{center}
 \vspace{-20pt}
 \caption{An overview of the proposed anomaly detection method. To obtain anomaly score, the pair of images reconstructed from the anomaly map and the anomaly detector (fake pair) are compared with the pair of the normal mask and the input image (real pair) using a discriminator.}
\label{fig5}
 \vspace{-10pt}
\end{wrapfigure}

\subsection{Anomaly Detection Using Proposed Anomaly Map}
In this section, we design a simple anomaly detector that adds the proposed anomaly map to the existing GAN-based detection method (\cite{alocc}). The proposed anomaly detector performs anomaly detection by learning only normal data distribution. We simply concatenate the input image and anomaly map to use them as inputs of detector, and apply both an adversarial loss and a reconstruction loss. Then, we use the feature matching loss introduced in (\cite{imp}) to stabilize the learning of the discriminator and extract the anomaly score. We include a detailed description of the training process for anomaly detection in Appendix A. 

In the test process (Fig. \ref{fig5}), the proposed anomaly detector obtains anomaly scores using the discriminator that has learned the normal data distribution. We first assume that the input image is normal, so the mask $A_{Nor}$ with all inner values set to one is used in pairs with the input image. When the input image is really normal, a fake pair (anomaly map and reconstructed image) is similar to the real pair (normal mask and input image), so the anomaly detector has a low anomaly score. On the other hand, when the input image is abnormal, the fake pair is significantly different to the real pair, so it has a high anomaly score. To compare the real and fake pair, the reconstruction loss and the feature matching loss are used as follows:
\begin{equation}
 Score = \alpha L_{MSE}(f(concat(x_{Seg},A_{Nor})), f(concat(\widehat{x}_{Seg},\widehat{A}_{Seg}))) + \beta L_{MSE}(x_{Seg}, \widehat{x}_{Seg}),
\end{equation}
where $\alpha$ and $\beta$ are 1 and 0.1, respectively. $A_{Nor}$ and $L_{MSE}$ represent a normal GT mask and the mean squared error, respectively.

\begin{table*}
\begin{center}
\label{table:headings}
\caption{Performance comparison of anomaly segmentation and detection in terms of pixel-level AUROC and image-level AUROC with the proposed method and conventional SOTA methods on the MVTec AD dataset (\cite{mvtec}). Full results for anomaly detection are added in Table 4 of Appendix A.3.}
\makeatletter
\def\hlinewd#1{
\noalign{\ifnum0=‘}\fi\hrule \@height #1 \futurelet
\reserved@a\@xhline}

\newcommand{\hthickline}{\hlinewd{1pt}}
\newcommand{\hthinline}{\hlinewd{.2pt}}
\makeatother
\newcolumntype{Z}{>{\centering\arraybackslash}X}
% {
% \footnotesize
\begin{tabularx}{\linewidth}{c||Z|Z|Z|Z|Z|Z|Z|Z}
\hthickline
  &\multicolumn{8}{c}{Anomaly Segmentation (Pixel-level AUROC)}\\\hline
\multirow{2}{*}{Method} &\multirow{2}{*}{AE$_{L2}$} &\text{\!\multirow{2}{*}{CAVGA}} &\multirow{2}{*}{US} &\multirow{2}{*}{FCDD} &Patch SVDD &\multirow{2}{*}{SPADE}  &\text{\!\!\multirow{2}{*}{Cutpaste} } &\text{\multirow{2}{*}{\!\!Proposed}}\\
%Method&AE_{L2} &CAVGA &US &FCDD &Patch SVDD &SPADE &Cutpaste &Proposed
\hline\noalign{\smallskip}
\hline
Bottle     & 0.86 & 0.89 & 0.94   & 0.97 & 0.98 & 0.98 & 0.98 & \textbf{0.99} \\\hline
Cable      & 0.86 & 0.85 &  0.91  & 0.90 & 0.97 & 0.97 & 0.90 & \textbf{0.99} \\\hline
Capsule    & 0.88 & 0.95 &  0.92  & 0.93 & 0.96 & \textbf{0.99} & 0.97 & 0.90 \\\hline
Carpet     & 0.59 & 0.88 &  0.72  & 0.96 & 0.93 & 0.98 & 0.98 & \textbf{0.99} \\\hline
Grid       & 0.90 & 0.95 &  0.85  & 0.91 & 0.96 & 0.94 & 0.98 & \textbf{0.99} \\\hline
Hazelnut   & 0.95 & 0.96 &  0.95  & 0.95 & 0.98 & \textbf{0.99} & 0.97 & \textbf{0.99} \\\hline
Leather    & 0.75 & 0.94 &   0.84 & 0.98 & 0.97 & 0.98 & \textbf{0.99} & 0.98 \\\hline
Metal\_nut  & 0.86 & 0.85 &   0.92 & 0.94 & 0.98 & 0.98 & 0.93 & \textbf{0.99} \\\hline
Pill       & 0.85 & 0.94 &   0.91 & 0.81 & 0.95 & \textbf{0.96} & \textbf{0.96} & 0.94 \\\hline
Screw      & 0.96 & 0.85 &  0.92  & 0.86 & 0.96 & \textbf{0.99} & 0.97 & 0.91 \\\hline
Tile       & 0.51 & 0.80 &  0.91  & 0.91 & 0.91 & 0.87 & 0.90 & \textbf{0.98} \\\hline
Toothbrush & 0.93 & 0.91 &  0.88  & 0.94 & \textbf{0.98} & \textbf{0.98} & \textbf{0.98} & 0.96 \\\hline
Transistor & 0.86 & 0.85 &  0.73  & 0.88 & \textbf{0.97} & 0.94 & 0.93 & 0.96 \\\hline
Wood       & 0.73 & 0.86 &  0.85  & 0.88 & 0.91 & 0.89 & 0.96 & \textbf{0.98} \\\hline
Zipper     & 0.77 & 0.94 & 0.91   & 0.92 & 0.95 & 0.97 & \textbf{0.99} & 0.98 \\\hline\hline
Mean       & 0.82 & 0.89 &  0.88  & 0.92 & 0.96 & 0.96 & 0.96 & \textbf{0.97}\\\hline
  &\multicolumn{8}{c}{Anomaly Detection (Image-level AUROC)}\\\hline
Mean &0.71 &0.82 &0.84 &- &0.92 &0.86 &0.95 &\textbf{0.96} \\\hline
\hthickline 
\end{tabularx}
%}
\end{center}
\vspace{-0.3cm}
\end{table*}

\section{Experimental Results}
\subsection{Evaluation Datasets and Metrics}
To verify the anomaly segmentation and detection performance of the proposed method, several evaluations were performed on the MVTec AD dataset (\cite{mvtec}). For the MVTec AD dataset, we resized both training and testing images to the size of 256 × 256, and each training batch contains 16 images. Following the previous works (\cite{mvtec,eatt, super}), we adopted the pixel-level and image-level AUROCs to quantitatively evaluate the performance of different methods for anomaly segmentation and detection, respectively. In addition, we used IoU to evaluate anomaly segmentation. For the measurement of IoU, a threshold, which maximizes IoU, was applied in each method.
 
\subsection{Implementation Details}
The encoder of AnoSeg consists of the convolution layers of ResNet-18 (\cite{res}). The up-sampling layer of decoders consists of one transposed convolution layer and convolution layers. Two decoders of the AnoSeg are composed of five up-sampling layers and two convolution layer to generate an anomaly map and a reconstructed image. The structure of the anomaly detector is the same as the AnoSeg structure except for the decoder that generates the anomaly map. Detailed information on training process and the network architecture is described in Appendix B.

%Detailed information on training process and network architectures is described in our supplementary material.

% \begin{figure*}[t]
% \begin{center}
% \includegraphics[width=1\linewidth]{fig4.png} 
%   %\includegraphics[width=0.8\linewidth]{egfigure.eps}
% \end{center}
%   \caption{Qualitative results of main experiments on MVtec AD dataset for (first row) input image, (second row) ground truth, and (third row) proposed anomaly map.}
   
% \label{fig5}
% \end{figure*}

\subsection{Experiments on the MVTec AD Dataset}

\subsubsection{Compared Methods}
We compared the reconstruction-based method with the proposed method using autoencoder-L2 ($\text{AE}_{L2}$). GradCAM-based methods (CAVGA (\cite{eatt}) and Cutpaste (\cite{cut})) were also compared with the proposed method. Also, we compared the proposed method with the US \cite{stu} using the representation of patch images. In our experiment, we compared the US trained with a patch size of $65\times65$. The proposed method is also compared with FCDD (\cite{fcdd}) using receptive field upsampling. Finally, among the embedding similarity-based methods, the patch SVDD (\cite{patch}) and SPADE (\cite{spa}) were also used for the performance comparison.

\begin{figure}[t]
\begin{center}
\includegraphics[width=1.0\linewidth]{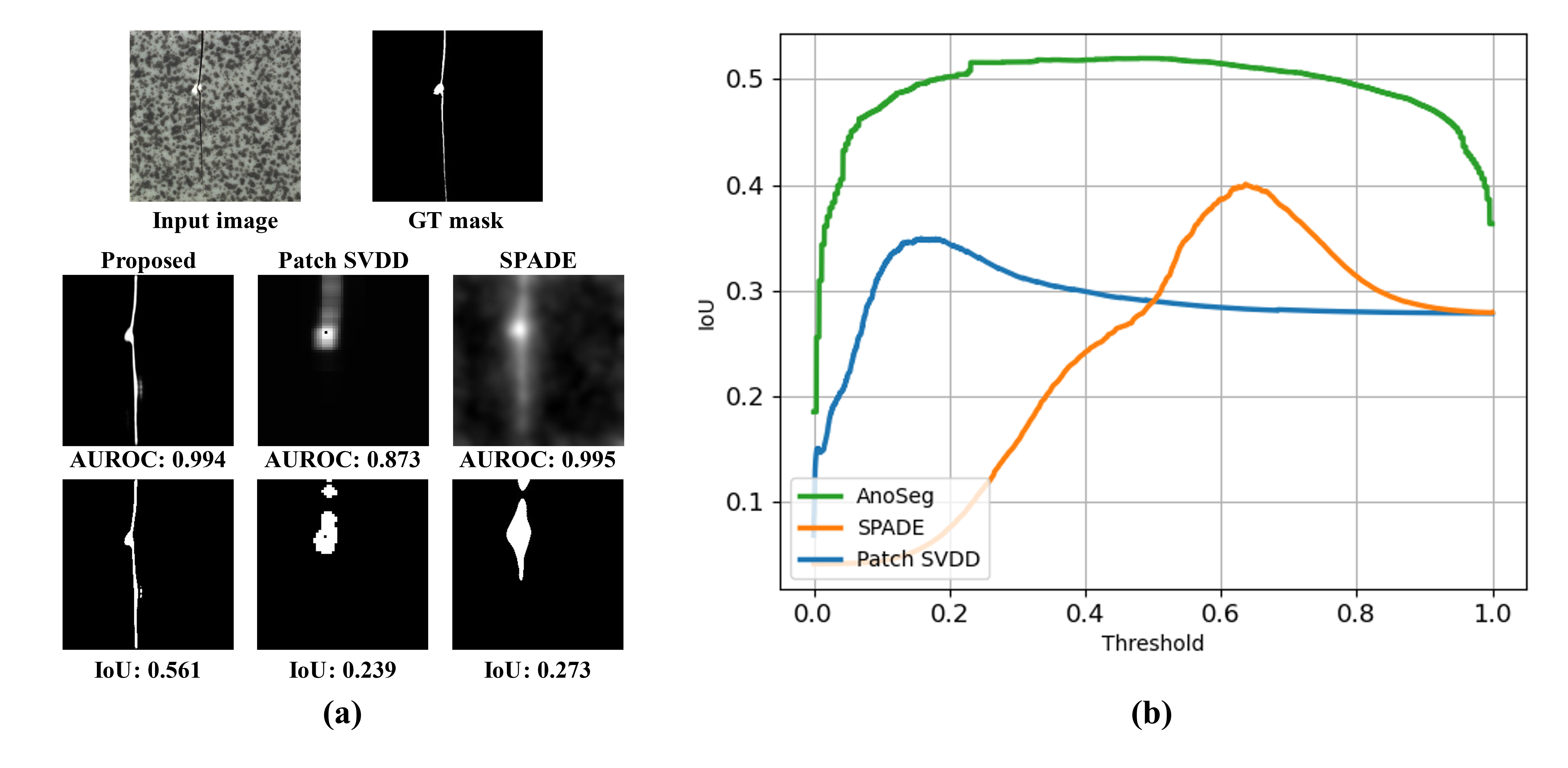} 
\end{center}
\vspace{-0.3cm}
  \caption{(a) Comparison on AUROC and IoU using Anomaly map and (b) mean IoU change according to the threshold for each category. The x-axis and y-axis represent a threshold and IoU, respectively.}
  \vspace{-0.3cm}
\label{fig6}

\end{figure}

\begin{table*}
\begin{center}
\label{table:headings}
\caption{Performance comparison of anomaly segmentation in term of mean IoU with the proposed and conventional SOTA methods on the MVTec AD dataset.}
\makeatletter
\def\hlinewd#1{%
\noalign{\ifnum0=‘}\fi\hrule \@height #1 \futurelet
\reserved@a\@xhline}
\newcommand{\hthickline}{\hlinewd{1pt}}
\newcommand{\hthinline}{\hlinewd{.2pt}}
\makeatother
\newcolumntype{Z}{>{\centering\arraybackslash}X}
{\footnotesize
\begin{tabularx}{\linewidth}{c||Z|Z|Z|Z|Z|Z}
\hthickline
  &\multicolumn{5}{c}{Anomaly Segmentation (IoU)}\\\hline
%Method &AnoGAN \cite{anog} &AE-SS \cite{mvtec} &AE-L2 \cite{mvtec} &VEVAE \cite{vavae} &CAVGA_D \cite{eatt} &Superpixel \cite{super} &Student \cite{stu} &Proposed &Semi\\
%\multirow{2}{*}{Method} &\multirow{2}{*}{AE_{L2}} &\multirow{2}{*}{CAVGA} &\multirow{2}{*}{US_{17}} &\multirow{2}{*}{US_{65}} &\multirow{2}{*}{FCDD} &\multirow{2}{*}{Patch SVDD} &Proposed \\
Method &CAVGA &US &Patch SVDD &SPADE &Proposed \\
\hline%\noalign{\smallskip}
Mean &0.470 &0.244 &0.427 &0.483 &\textbf{0.542} \\\hline
\hthickline 
\end{tabularx}
}
\vspace{-0.3cm}
\end{center}
\end{table*}

\subsubsection{Quantitative Results}
We evaluated the anomaly segmentation performance between the proposed method and the existing SOTA methods mentioned in section 4.3.1 using the MVTec AD dataset. As shown in Table 1, the proposed method consistently outperformed all other existing methods evaluated in AUROC. The reconstruction-based methods such as $\text{AE}_{L2}$ used the reconstruction loss as the anomaly score. $\text{AE}_{L2}$ had lower performance (0.82 AUROC) compared to the proposed method. CAVGA (\cite{eatt}) and Cutpaste (\cite{cut}) obtained anomaly maps using GradCAM (\cite{grad}), but these anomaly maps highly depend on the classification loss. In addition, compared to the methods using patch image representation such as US, the proposed method achieved higher performance. As a result, AnoSeg outperformed the conventional SOTA, such as Patch SVDD, SPADE, and Cutpaste, by $1\%$ AUROC in anomaly segmentation.

In addition, we evaluated IoU, which is typically used as a metric for segmentation. Table 2 shows the quantitative comparison on IoU. AnoSeg achieved the highest performance compared to other methods in IoU. In particular, Patch SVDD and SPADE achieved 0.96 AUROC similar to AnoSeg in the evaluation of AUROC, but had lower IoU than the proposed method. This is because, unlike the existing method, the proposed method was directly trained for segmentation.

Additionally, we compared the AUROC and IoU metrics for the generated anomaly map in Fig. \ref{fig6}(a). In general, AUROC is affected by the detection performance of the anomaly regions. False positives for normal regions have relatively no impact on AUROC. In the Patch SVDD of Fig. \ref{fig6}(a), there were abnormal regions that cannot be detected. Therefore, the anomaly map of Patch SVDD had lower AUROC compared to other methods. Although the anomaly maps of AnoSeg and SPADE visually show different anomaly maps, the same AUROC was calculated because most anomaly regions are detected in anomaly maps of AnoSeg and SPADE. However, IoU was affected by false positives in normal regions. Therefore, IoU of SPADE had lower performance compared to AUROC. The proposed AnoSeg achieved the highest performance for both IoU and AUROC. These results shows that the proposed method is superior in various aspects of anomaly segmentation. 

We compared the anomaly detection performance between the proposed and existing methods using the method introduced in section 4.3.1. As shown in Table 1, the proposed method achieved similar AUROC to existing SOTA methods (Full results are in Appendix A.3). Discriminator of anomaly detector learned representations of images and anomaly maps together. Therefore, with a simple anomaly detection method using the generated anomaly map, we achieve anomaly detection performance similar to that of the existing SOTA.

\begin{figure}[t]
\begin{center}
\includegraphics[width=1.0\linewidth]{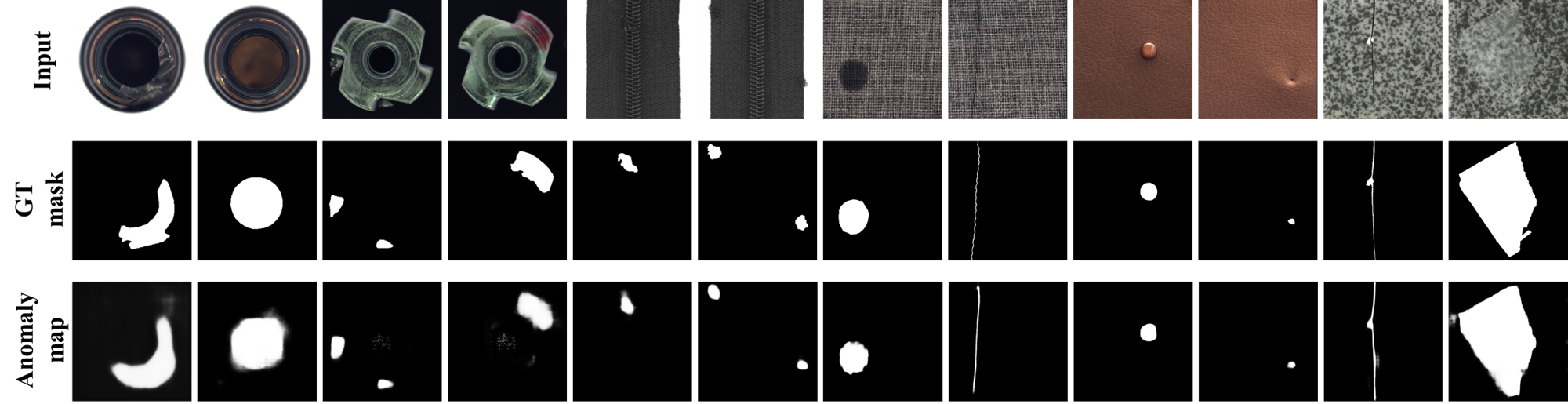} 
\end{center}
   \caption{Qualitative results on the MVTec AD dataset for (first row) input image, (second row) GT mask, and (third row) proposed anomaly map.}
\label{fig7}
%\vspace{-0.4cm}
\end{figure}

\begin{table*}
\begin{center}
\label{table:headings}
\caption{Performance of various configurations on the MVTec AD dataset.}
\makeatletter
\def\hlinewd#1{%
\noalign{\ifnum0=‘}\fi\hrule \@height #1 \futurelet
\reserved@a\@xhline}
\newcommand{\hthickline}{\hlinewd{1pt}}
\newcommand{\hthinline}{\hlinewd{.2pt}}
\makeatother
\newcolumntype{Z}{>{\centering\arraybackslash}X}
{\footnotesize
\begin{tabularx}{\linewidth}{c||Z|Z|Z|Z}
\hthickline
  &\multicolumn{4}{c}{Ablation study (AUROC / IoU)}\\\hline
%Method &AnoGAN \cite{anog} &AE-SS \cite{mvtec} &AE-L2 \cite{mvtec} &VEVAE \cite{vavae} &CAVGA_D \cite{eatt} &Superpixel \cite{super} &Student \cite{stu} &Proposed &Semi\\
%\multirow{2}{*}{Method} &\multirow{2}{*}{AE_{L2}} &\multirow{2}{*}{CAVGA} &\multirow{2}{*}{US_{17}} &\multirow{2}{*}{US_{65}} &\multirow{2}{*}{FCDD} &\multirow{2}{*}{Patch SVDD} &Proposed \\
Method &Base model (Cutpaste only) & + Hard augmentation & + Adversarial learning & + Coordinate channel \\
\hline%\noalign{\smallskip}
Mean &0.923 / 0.492 &0.942 / 0.503 &0.951 / 0.527 &0.970 / 0.542\\\hline
\hthickline 
\end{tabularx}
}
\vspace{-0.3cm}
\end{center}
\end{table*}

\subsubsection{Qualitative Results}

For the evaluation with existing methods, we visualized anomaly maps of existing and proposed methods in Fig. \ref{fig1}. The output image of $\text{AE}_{L2}$ (\cite{mvtec}) was restored up to the anomaly image region and it was difficult to restore high-frequency regions of the normal image. Also, $\text{US}_{65\times65}$ could detect large defects, but had poor detection performance for small defects. These results show that patch representations based methods are difficult to accurately localize defects for various sizes. Patch SVDD and SPADE extracted anomaly maps using feature extractions for different sizes to consider defects with various sizes. Therefore, the defects with different sizes could be detected, as shown in Fig. \ref{fig1}. However, these anomaly maps had many false positives for normal regions and approximately detected anomaly regions. In contrast, as shown in Fig. \ref{fig7}, the proposed AnoSeg was trained to generate anomaly maps directly for anomaly segmentation using the segmentation loss. Therefore, the proposed method generated an anomaly map more similar to GT than the results of the existing methods as shown in Fig. 6. More comprehensive results on defect segmentation are given in Appendix C.

\subsubsection{Analysis of Threshold Sensitivity}
In this section, Patch SVDD and our AnoSeg were compared to verify the performance variation depending on the threshold of the proposed method. IoU was measured by dividing the anomaly score by 10000 units. Fig. \ref{fig6}(b) shows the performance change of AnoSeg, SPADE and Patch SVDD according to a threshold. As shown in Fig. \ref{fig6}(b), the performance of AnoSeg did not significantly change significantly for different thresholds. Therefore, the anomaly map is shown similar to the GT mask even though thresholding was not applied in Fig. \ref{fig6}. On the other hand, Fig. \ref{fig6}(b) shows that Patch SVDD and SPADE had a significant change in performance when the threshold is changed around the threshold with the highest IoU. The result shows that our model is robust against thresholding. By setting the threshold between 0.2 and 0.8, AnoSeg could always achieve better results consistently than other SOTA solutions listed in Table 2.

\section{Ablation Study} 
We modified the generator structure (Section 4.2) to generate the only anomaly map and construct the base model with only Cutpaste applied. Then, we added modules incrementally on the base model, and evaluated with IoU and AUROC scores. The overall results show that the method using all modules improved by 5.4\% and 10.2\% for AUROC and IoU, respectively, compared to the base model. The effectiveness of each module is described below.\\

\begin{figure}[t]
\begin{center}
\includegraphics[width=0.9\linewidth]{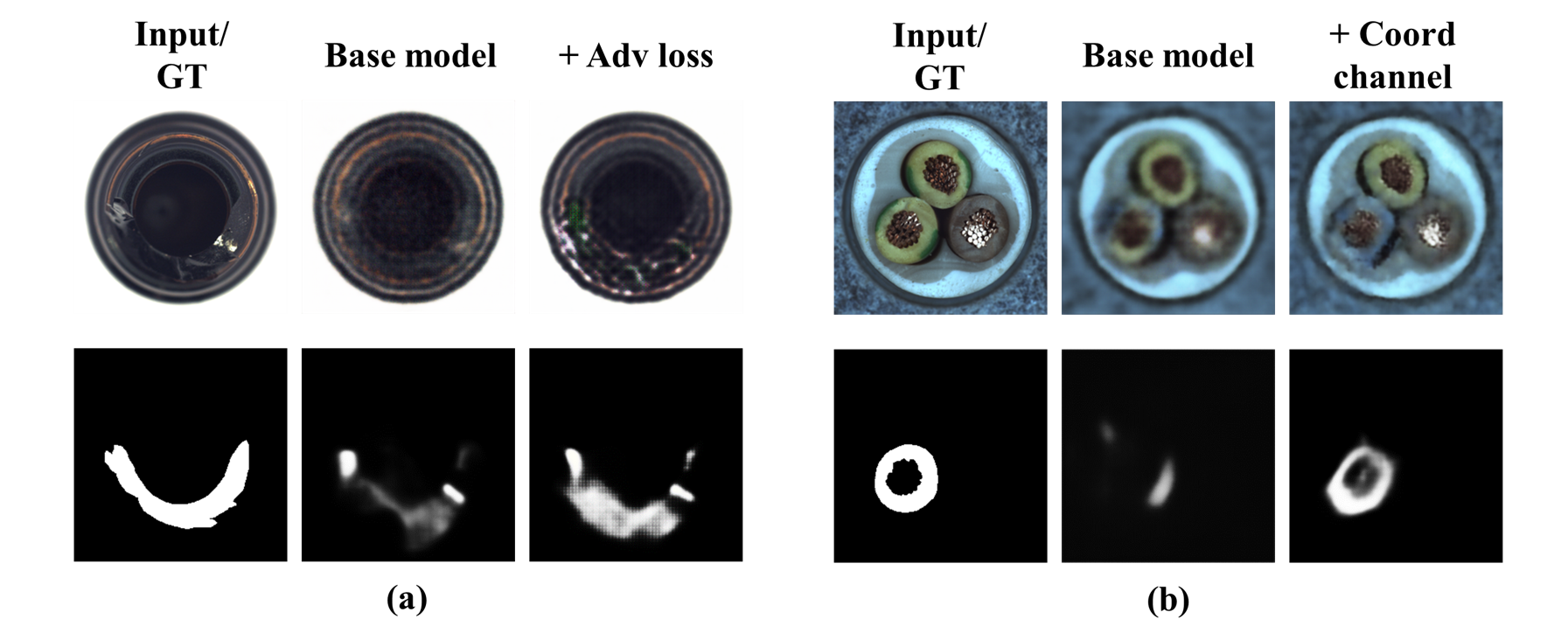} 

\end{center}
\vspace{-0.2cm}
\caption{Qualitative results of the ablation study to illustrate the performance of the anomaly segmentation on the MVtec AD dataset.}
\vspace{-0.2cm}
\label{fig3}
\end{figure}

\textbf{Hard augmentation} \quad We used images with several hard augmentations applied to train AnoSeg on anomaly regions. Hard augmentations generate samples away from the normal data distribution. Intuitively, synthetic anomaly data applied with hard augmentation can generate more diverse anomaly regions than Cutpaste. Therefore, AnoSeg detected more anomaly regions than the base model. As a result, AUROC and IOU were improved by 2.1\% and 1.9\% respectively.

\textbf{Adversarial learning with reconstruction loss} \quad The proposed AnoSeg learns the normal region distribution through adversarial learning. We also use masked reconstruction loss in AnoSeg to apply reconstruction loss only for normal regions to avoid biasing synthetic anomaly regions. As shown in a of Fig. 8(a), the base model is difficult to learn the normal data distribution. Therefore, the reconstructed image of base model partially restores the anomaly regions, and the base model detects anomaly regions as normal regions. In contrast, a model using adversarial learning learns the normal data distribution and can segment between normal and abnormal regions. Therefore, AnoSeg can generate detailed anomaly maps.

\textbf{Coordinate channel concatenation} \quad To consider the additional location information while performing anomaly segmentation, we concatenated coordinate channels. In Fig. 8(b), the effectiveness of coordinate channel concatenation is confirmed. The yellow cable in the input image changes the class property depending on the location. Therefore, these anomaly regions can be determine as normal if location information is insufficient. Because the base model that does not use the coordinate channel lacks location information, the yellow cable, which is an abnormal area, is reconstructed and determined as a normal area. AnoSeg provides additional location information by connecting the coordinate channel to the input image. As a result, as shown in Fig.8(b) anomaly regions that depend on location information were additionally detected, and AUROC and IOU were improved by 1.9\% and 2.8\% respectively.

\section{Conclusion}
This paper presented a novel anomaly segmentation network to directly generate an anomaly map. We proposed AnoSeg, a segmentation model using adversarial learning, and the proposed AnoSeg was directly trained for anomaly segmentation using synthetic anomaly data generated through hard augmentation. In addition, anomaly regions sensitive to positional relationships were more easily detected through coordinate vectors representing the pixel position information. Hence, our approach enabled AnoSeg to be trained to generate anomaly maps with direct supervision. We also applied this anomaly maps to existing methods to improve the performance of anomaly detection. Experimental results on the MVTec AD dataset using AUROC and IoU demonstrated that the proposed method is a specialized network for anomaly segmentation compared to the existing methods.
%This paper presented a novel hard augmentation-based two-stage framework consisting of the ASGAN and ADGAN to directly generate anomaly map representing the normal region and easily detect the anomaly class. We propose ASGAN, a segmentation model using adversarial learning, and the proposed ASGAN is directly trained for anomaly segmentation using synthetic anomaly data generated through hard augmentation. Hence, our approach enabled ASGAN to be trained to generate anomaly map with direct supervision. Moreover, through this anomaly map, ADGAN learned the normal class distribution of normal image and anomaly map. Since the Anomaly map of the anomaly class was generated differently from the normal class distribution, Anomaly images are detected more easily than conventional methods. Experimental results show that the proposed method outperformed the existing SOTA anomaly detection and segmentation methods on mvtec ad datasets.

\bibliography{iclr2022_conference.bbl}

\begin{thebibliography}{26}
\providecommand{\natexlab}[1]{#1}
\providecommand{\url}[1]{\texttt{#1}}
\expandafter\ifx\csname urlstyle\endcsname\relax
  \providecommand{\doi}[1]{doi: #1}\else
  \providecommand{\doi}{doi: \begingroup \urlstyle{rm}\Url}\fi

\bibitem[Akcay et~al.(2018)Akcay, Atapour-Abarghouei, and Breckon]{re7}
Samet Akcay, Amir Atapour-Abarghouei, and Toby~P. Breckon.
\newblock Ganomaly: Semi-supervised anomaly detection via adversarial training,
  2018.

\bibitem[An \& Cho(2015)An and Cho]{re8}
Jinwon An and Sungzoon Cho.
\newblock Variational autoencoder based anomaly detection using reconstruction
  probability.
\newblock \emph{Special Lecture on IE}, 2\penalty0 (1):\penalty0 1--18, 2015.

\bibitem[Baur et~al.(2018)Baur, Wiestler, Albarqouni, and Navab]{re9}
Christoph Baur, Benedikt Wiestler, Shadi Albarqouni, and Nassir Navab.
\newblock Deep autoencoding models for unsupervised anomaly segmentation in
  brain mr images.
\newblock In \emph{International MICCAI Brainlesion Workshop}, pp.\  161--169.
  Springer, 2018.

\bibitem[{Bergmann} et~al.(2019){Bergmann}, {Fauser}, {Sattlegger}, and
  {Steger}]{mvtec}
P.~{Bergmann}, M.~{Fauser}, D.~{Sattlegger}, and C.~{Steger}.
\newblock Mvtec ad — a comprehensive real-world dataset for unsupervised
  anomaly detection.
\newblock In \emph{2019 IEEE/CVF Conference on Computer Vision and Pattern
  Recognition (CVPR)}, pp.\  9584--9592, 2019.

\bibitem[Bergmann et~al.(2020)Bergmann, Fauser, Sattlegger, and Steger]{stu}
Paul Bergmann, Michael Fauser, David Sattlegger, and Carsten Steger.
\newblock Uninformed students: Student-teacher anomaly detection with
  discriminative latent embeddings.
\newblock In \emph{Proceedings of the IEEE/CVF Conference on Computer Vision
  and Pattern Recognition}, pp.\  4183--4192, 2020.

\bibitem[Chen et~al.(2017)Chen, Sathe, Aggarwal, and Turaga]{re10}
Jinghui Chen, Saket Sathe, Charu Aggarwal, and Deepak Turaga.
\newblock Outlier detection with autoencoder ensembles.
\newblock In \emph{Proceedings of the 2017 SIAM international conference on
  data mining}, pp.\  90--98. SIAM, 2017.

\bibitem[Cohen \& Hoshen(2020)Cohen and Hoshen]{spa}
Niv Cohen and Yedid Hoshen.
\newblock Sub-image anomaly detection with deep pyramid correspondences.
\newblock \emph{arXiv preprint arXiv:2005.02357}, 2020.

\bibitem[Deecke et~al.(2018)Deecke, Vandermeulen, Ruff, Mandt, and Kloft]{re11}
Lucas Deecke, Robert Vandermeulen, Lukas Ruff, Stephan Mandt, and Marius Kloft.
\newblock Image anomaly detection with generative adversarial networks.
\newblock In \emph{Joint european conference on machine learning and knowledge
  discovery in databases}, pp.\  3--17. Springer, 2018.

\bibitem[Goodfellow et~al.(2014)Goodfellow, Pouget-Abadie, Mirza, Xu,
  Warde-Farley, Ozair, Courville, and Bengio]{gan}
Ian Goodfellow, Jean Pouget-Abadie, Mehdi Mirza, Bing Xu, David Warde-Farley,
  Sherjil Ozair, Aaron Courville, and Yoshua Bengio.
\newblock Generative adversarial nets.
\newblock In \emph{Advances in neural information processing systems}, pp.\
  2672--2680, 2014.

\bibitem[He et~al.(2016)He, Zhang, Ren, and Sun]{res}
Kaiming He, Xiangyu Zhang, Shaoqing Ren, and Jian Sun.
\newblock Deep residual learning for image recognition.
\newblock In \emph{Proceedings of the IEEE conference on computer vision and
  pattern recognition}, pp.\  770--778, 2016.

\bibitem[Kimura et~al.(2020)Kimura, Chaudhury, Narita, Munawar, and
  Tachibana]{att}
Daiki Kimura, Subhajit Chaudhury, Minori Narita, Asim Munawar, and Ryuki
  Tachibana.
\newblock Adversarial discriminative attention for robust anomaly detection.
\newblock In \emph{The IEEE Winter Conference on Applications of Computer
  Vision}, pp.\  2172--2181, 2020.

\bibitem[Kingma \& Welling(2014)Kingma and Welling]{vae}
Diederik Kingma and Max Welling.
\newblock Auto-encoding variational bayes.
\newblock In \emph{International Conference on Learning Representations}, 12
  2014.

\bibitem[Li et~al.(2021)Li, Sohn, Yoon, and Pfister]{cut}
Chun-Liang Li, Kihyuk Sohn, Jinsung Yoon, and Tomas Pfister.
\newblock Cutpaste: Self-supervised learning for anomaly detection and
  localization.
\newblock In \emph{Proceedings of the IEEE/CVF Conference on Computer Vision
  and Pattern Recognition}, pp.\  9664--9674, 2021.

\bibitem[Li et~al.(2020)Li, Li, Jiang, Ma, Wei, Hong, and Gong]{super}
Zhenyu Li, Ning Li, Kaitao Jiang, Zhiheng Ma, Xing Wei, Xiaopeng Hong, and
  Yihong Gong.
\newblock Superpixel masking and inpainting for self-supervised anomaly
  detection.
\newblock In \emph{31st British Machine Vision Conference 2020, {BMVC} 2020,
  Virtual Event, UK, September 7-10, 2020}. {BMVA} Press, 2020.
\newblock URL \url{https://www.bmvc2020-conference.com/assets/papers/0275.pdf}.

\bibitem[Liu et~al.(2018)Liu, Lehman, Molino, Petroski~Such, Frank, Sergeev,
  and Yosinski]{coord}
Rosanne Liu, Joel Lehman, Piero Molino, Felipe Petroski~Such, Eric Frank, Alex
  Sergeev, and Jason Yosinski.
\newblock An intriguing failing of convolutional neural networks and the
  coordconv solution.
\newblock In S.~Bengio, H.~Wallach, H.~Larochelle, K.~Grauman, N.~Cesa-Bianchi,
  and R.~Garnett (eds.), \emph{Advances in Neural Information Processing
  Systems}, volume~31. Curran Associates, Inc., 2018.
\newblock URL
  \url{https://proceedings.neurips.cc/paper/2018/file/60106888f8977b71e1f15db7bc9a88d1-Paper.pdf}.

\bibitem[Liznerski et~al.(2021)Liznerski, Ruff, Vandermeulen, Franks, Kloft,
  and Muller]{fcdd}
Philipp Liznerski, Lukas Ruff, Robert~A. Vandermeulen, Billy~Joe Franks, Marius
  Kloft, and Klaus~Robert Muller.
\newblock Explainable deep one-class classification.
\newblock In \emph{International Conference on Learning Representations}, 2021.
\newblock URL \url{https://openreview.net/forum?id=A5VV3UyIQz}.

\bibitem[Odena(2016)]{semi}
Augustus Odena.
\newblock Semi-supervised learning with generative adversarial networks.
\newblock \emph{arXiv preprint arXiv:1606.01583}, 2016.

\bibitem[Sabokrou et~al.(2018)Sabokrou, Khalooei, Fathy, and Adeli]{alocc}
Mohammad Sabokrou, Mohammad Khalooei, Mahmood Fathy, and Ehsan Adeli.
\newblock Adversarially learned one-class classifier for novelty detection.
\newblock In \emph{Proceedings of the IEEE Conference on Computer Vision and
  Pattern Recognition}, pp.\  3379--3388, 2018.

\bibitem[Sakurada \& Yairi(2014)Sakurada and Yairi]{re12}
Mayu Sakurada and Takehisa Yairi.
\newblock Anomaly detection using autoencoders with nonlinear dimensionality
  reduction.
\newblock In \emph{Proceedings of the MLSDA 2014 2nd Workshop on Machine
  Learning for Sensory Data Analysis}, pp.\  4--11, 2014.

\bibitem[Salimans et~al.(2016)Salimans, Goodfellow, Zaremba, Cheung, Radford,
  Chen, and Chen]{imp}
Tim Salimans, Ian Goodfellow, Wojciech Zaremba, Vicki Cheung, Alec Radford,
  Xi~Chen, and Xi~Chen.
\newblock Improved techniques for training gans.
\newblock In D.~Lee, M.~Sugiyama, U.~Luxburg, I.~Guyon, and R.~Garnett (eds.),
  \emph{Advances in Neural Information Processing Systems}, volume~29. Curran
  Associates, Inc., 2016.
\newblock URL
  \url{https://proceedings.neurips.cc/paper/2016/file/8a3363abe792db2d8761d6403605aeb7-Paper.pdf}.

\bibitem[Schlegl et~al.(2017)Schlegl, Seeb{\"o}ck, Waldstein, Schmidt-Erfurth,
  and Langs]{anog}
Thomas Schlegl, Philipp Seeb{\"o}ck, Sebastian~M Waldstein, Ursula
  Schmidt-Erfurth, and Georg Langs.
\newblock Unsupervised anomaly detection with generative adversarial networks
  to guide marker discovery.
\newblock In \emph{International conference on information processing in
  medical imaging}, pp.\  146--157. Springer, 2017.

\bibitem[Selvaraju et~al.(2017)Selvaraju, Cogswell, Das, Vedantam, Parikh, and
  Batra]{grad}
Ramprasaath~R Selvaraju, Michael Cogswell, Abhishek Das, Ramakrishna Vedantam,
  Devi Parikh, and Dhruv Batra.
\newblock Grad-cam: Visual explanations from deep networks via gradient-based
  localization.
\newblock In \emph{Proceedings of the IEEE international conference on computer
  vision}, pp.\  618--626, 2017.

\bibitem[Tack et~al.(2020)Tack, Mo, Jeong, and Shin]{csi}
Jihoon Tack, Sangwoo Mo, Jongheon Jeong, and Jinwoo Shin.
\newblock Csi: Novelty detection via contrastive learning on distributionally
  shifted instances.
\newblock In H.~Larochelle, M.~Ranzato, R.~Hadsell, M.~F. Balcan, and H.~Lin
  (eds.), \emph{Advances in Neural Information Processing Systems}, volume~33,
  pp.\  11839--11852. Curran Associates, Inc., 2020.
\newblock URL
  \url{https://proceedings.neurips.cc/paper/2020/file/8965f76632d7672e7d3cf29c87ecaa0c-Paper.pdf}.

\bibitem[Venkataramanan et~al.(2020)Venkataramanan, Peng, Singh, and
  Mahalanobis]{eatt}
Shashanka Venkataramanan, Kuan-Chuan Peng, Rajat~Vikram Singh, and Abhijit
  Mahalanobis.
\newblock Attention guided anomaly localization in images.
\newblock In \emph{Proceedings of the European Conference on Computer Vision
  (ECCV)}, September 2020.

\bibitem[Yi \& Yoon(2020)Yi and Yoon]{patch}
Jihun Yi and Sungroh Yoon.
\newblock Patch svdd: Patch-level svdd for anomaly detection and segmentation.
\newblock In \emph{Proceedings of the Asian Conference on Computer Vision},
  2020.

\bibitem[Zenati et~al.(2018)Zenati, Foo, Lecouat, Manek, and
  Chandrasekhar]{re14}
Houssam Zenati, Chuan~Sheng Foo, Bruno Lecouat, Gaurav Manek, and
  Vijay~Ramaseshan Chandrasekhar.
\newblock Efficient gan-based anomaly detection.
\newblock \emph{arXiv preprint arXiv:1802.06222}, 2018.

\end{thebibliography}
\bibliographystyle{iclr2022_conference}

\appendix
\section{Anomaly Detection Using Proposed Anomaly Map}
Here we provide detailed information for the training and loss functions of anomaly detector using the proposed anomaly map from Section 3.4.
\subsection{Training Process of Anomaly Detection Method}
The proposed anomaly detection method uses an anomaly map generated from the AnoSeg along with the input image to learn the distribution of the normal image and the anomaly map. Therefore, the anomaly detector determines whether the anomaly map is focusing on the normal region of the input image while determining whether the input image is a normal image. Unlike AnoSeg, the proposed anomaly detection method does not use the synthetic anomaly $x_{Ano}$ as a real class in an adversarial loss because discriminator of anomaly detector only needs to learn the normal data distribution for anomaly detection. The loss function for learning the discriminator of the anomaly detector ($L_{Adv}^{AD}$) is as follows:
\begin{align}
  L_{Adv}^{AD} = \underset{G}{min} \underset{D}{max}\{\mathbb{E}\;[\log(1-D(concat(\widehat{x}_{Nor}, \widehat{A}_{Nor})))] \nonumber \\   +\mathbb{E}\;[\log(D(concat(x_{Nor},A_{Nor})))]\},
  \label{equ:dt}
\end{align}
where $\widehat{x}_{Nor}$, $\widehat{A}_{Nor}$, $x_{Nor}$, and $,A_{Nor}$ represent reconstructed a normal image, a anomaly map of AnoSeg, a normal image, and a normal mask, repectively.

Also, to help estimate the normal data distribution, we propose a synthetic anomaly classification loss that discriminates between synthetic data and normal data. As confirmed in (\cite{semi}), the proposed synthetic anomaly classification loss improves the anomaly performance of the discriminator. This synthetic anomaly classification loss is defined as:
\begin{align}
  L_{cls} = \mathbb{E}\;[\log(1-D(concat(x_{Ano},A_{Ano})))] \nonumber +\mathbb{E}\;[\log(D(concat(x_{Nor},A_{Nor})))].
  \label{equ:dt}
\end{align}

Then, we use the feature matching loss introduced in (\cite{imp}) to stabilize the learning of the discriminator and extract the anomaly score. The high-level representations of the normal and reconstructed samples are expected to be identical. This loss is given as follows:
\begin{align}
  L_{fea} = \mathbb{E}\parallel f(concat(x_{Nor},A_{Nor})) \nonumber -\,f(concat(\widehat{x}_{Nor},\widehat{A}_{Nor}))\parallel ^{2},
\end{align}
where $f(.)$ is the second to the last layer of the discriminator. Fig. 9 shows an overview of the overall training process.
\begin{figure}[t]
\begin{center}
\includegraphics[width=1.0\linewidth]{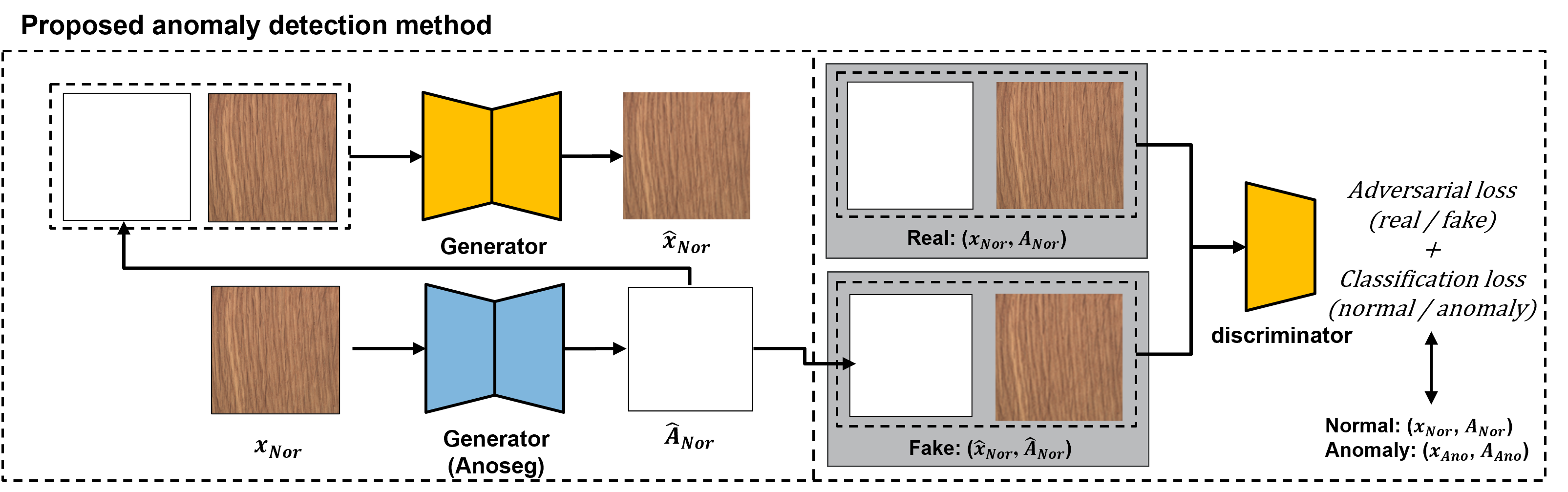} 
\end{center}
\vspace{-0.2cm}
  \caption{Overview of the training process of the proposed anomaly detection method.}
\label{fig2}
\end{figure}

\subsection{Quantitative Evaluation of Anomaly Detection in the MVTec AD dataset.}

We describe the performance evaluation setting of the existing method that was not included in the main paper due to the length limitation. For performance comparison with existing methods, we used the results from existing literature, excluding the uninformed students method (US) (\cite{stu}). US method is only evaluated with PRO scores for anomaly segmentation without the provision of the AUROC for the anomaly segmentation and detection. Therefore, we re-implemented the large patch size (patch size is $65 \times 65$) version of the Student method and evaluated it on anomaly detection and segmentation. Tables 4 also shows the class-wise anomaly detection performances for the MVTec AD (AUROC) dataset.

\begin{table*}
\begin{center}
\label{table:headings}
\caption{Performance comparison of anomaly detection in terms of image-level AUROC with the proposed method and conventional SOTA methods on the MVTec AD dataset (\cite{mvtec}).}
\makeatletter
\def\hlinewd#1{
\noalign{\ifnum0=‘}\fi\hrule \@height #1 \futurelet
\reserved@a\@xhline}
\newcommand{\hthickline}{\hlinewd{1pt}}
\newcommand{\hthinline}{\hlinewd{.2pt}}
\makeatother
\newcolumntype{Z}{>{\centering\arraybackslash}X}
{\footnotesize
\begin{tabularx}{\linewidth}{c||Z|Z|Z|Z|Z|Z|Z|Z}
\hthickline
  &\multicolumn{7}{c}{Anomaly Detection (Image-level AUROC)}\\\hline
%Method &AnoGAN \cite{anog} &AE-SS \cite{mvtec} &AE-L2 \cite{mvtec} &VEVAE \cite{vavae} &CAVGA_D \cite{eatt} &Superpixel \cite{super} &Student \cite{stu} &Proposed &Semi\\
%\multirow{2}{*}{Method} &\multirow{2}{*}{AE_{L2}} &\multirow{2}{*}{CAVGA} &\multirow{2}{*}{US_{17}} &\multirow{2}{*}{US_{65}} &\multirow{2}{*}{FCDD} &\multirow{2}{*}{Patch SVDD} &Proposed \\
\multirow{2}{*}{Method} &\multirow{2}{*}{AE$_{L2}$} &\!\multirow{2}{*}{CAVGA} &\multirow{2}{*}{US}  &Patch SVDD &\multirow{2}{*}{SPADE}  &\!\!\multirow{2}{*}{Cutpaste}  &\!\!\multirow{2}{*}{Proposed} \\
\hline\noalign{\smallskip}
\hline
bottle     & 0.80 & 0.91 &  0.85   & \textbf{0.99} & - &0.98 &0.98 \\\hline
Cable      & 0.56 & 0.67 &  0.90   & 0.90 & - & 0.81 & \textbf{0.98} \\\hline
Capsule    & 0.62 & 0.87 &  0.82   & 0.77 & - & \textbf{0.96} & 0.84 \\\hline
Carpet     & 0.50 & 0.78 &  0.86   & 0.93 & - & 0.93 & \textbf{0.96} \\\hline
Grid       & 0.78 & 0.78 &  0.60   & 0.95 & - &\textbf{0.99} & \textbf{0.99} \\\hline
Hazelnut   & 0.88 & 0.87 &  0.91   & 0.92 & - & 0.97 & \textbf{0.98} \\\hline
Leather    & 0.44 & 0.75 &  0.73   & 0.91 & - &\textbf{1.00} & 0.99 \\\hline
Metal\_nut & 0.73 & 0.71 &  0.58   & 0.94 & - & \textbf{0.99} & 0.95 \\\hline
Pill       & 0.62 & 0.91 &  0.90   & 0.86 &- & \textbf{0.92} & 0.87 \\\hline
Screw      & 0.69 & 0.78 &  0.90   & 0.81 & - & 0.86 & \textbf{0.97} \\\hline
Tile       & 0.77 & 0.72 &  0.87   & \textbf{0.98} & - & 0.93 & \textbf{0.98} \\\hline
Toothbrush & 0.98 & 0.97 &  0.81   & \textbf{1.00} & - & 0.98 & 0.99 \\\hline
Transistor & 0.71 & 0.75 &  0.85   & 0.92 & - & \textbf{0.96} & \textbf{0.96} \\\hline
Wood       & 0.74 & 0.88 &  0.68   & 0.92 & - &\textbf{0.99} & \textbf{0.99} \\\hline
Zipper     & 0.80 & 0.94 &  0.90   & 0.98 & - & \textbf{0.99} & \textbf{0.99} \\\hline\hline
Mean       & 0.71 & 0.82 &  0.84   & 0.92 & 0.86 & 0.95 & \textbf{0.96}\\\hline

\hthickline 
\end{tabularx}
}
\end{center}
\vspace{-0.2cm}
\end{table*}

\subsection{Ablation study of Anomaly Detection Method}
We evaluated the effectiveness of the individual components in the proposed anomaly detection method on the MVTec AD dataset, as shown in Table 5. The base model used the same structure as the proposed model, and only the input images were fed except for the mask. The base model compared the features of the input image and the reconstructed image to calculate an anomaly score. However, since the reconstructed image often had anomaly regions restored, the base model has the low performance. The model that the feature matching loss is applied had slightly improved AUROC than the base model. The proposed anomaly detection method performed anomaly detection using input images and anomaly maps. Image-level AUROC was significantly increased by up to 15\%. Hence, the model using an anomaly map as an input performed anomaly detection more sensitive than the conventional method using only an input image. Finally, to enhance the estimation of the normal data distribution, we added an anomaly classification loss. This loss helps in estimating the boundaries of the normal data distribution where synthetic anomaly data are separated. 

\begin{table*}
\begin{center}
\label{table:headings}
\caption{Anomaly detection performance of various configurations on the MVTec AD dataset.}
\makeatletter
\def\hlinewd#1{%
\noalign{\ifnum0=‘}\fi\hrule \@height #1 \futurelet
\reserved@a\@xhline}
\newcommand{\hthickline}{\hlinewd{1pt}}
\newcommand{\hthinline}{\hlinewd{.2pt}}
\makeatother
\newcolumntype{Z}{>{\centering\arraybackslash}X}
{\footnotesize
\begin{tabularx}{\linewidth}{c||Z|Z|Z|Z}
\hthickline
  &\multicolumn{4}{c}{Ablation study (Image-level AUROC)}\\\hline
%Method &AnoGAN \cite{anog} &AE-SS \cite{mvtec} &AE-L2 \cite{mvtec} &VEVAE \cite{vavae} &CAVGA_D \cite{eatt} &Superpixel \cite{super} &Student \cite{stu} &Proposed &Semi\\
%\multirow{2}{*}{Method} &\multirow{2}{*}{AE_{L2}} &\multirow{2}{*}{CAVGA} &\multirow{2}{*}{US_{17}} &\multirow{2}{*}{US_{65}} &\multirow{2}{*}{FCDD} &\multirow{2}{*}{Patch SVDD} &Proposed \\
\multirow{2}{*}{Method} & \multirow{2}{*}{Base model} & + Feature matching loss & + Input anomaly map & + Anomaly classification loss\\
\hline%\noalign{\smallskip}
Mean &0.812 &0.842 &0.943 &0.961\\\hline
\hthickline 
\end{tabularx}
}
\end{center}
\end{table*}

\section{Details on the Network Architectures}

Table 6 shows the network structure of the proposed method. Each network is described by a list of layers including an output shape, a kernel size, a padding size, and a stride. In addition, batch normalization (BN) and activation function define whether BN is applied and which activation function is applied, respectively. The decoder used for image reconstruction has the same structure as the decoder for generating anomaly map, and AnoSeg uses two decoders. The structure of the proposed anomaly detector also has the same structure as that of AnoSeg. The structure of the AnnoSeg is also available in our code added in the supplementary material. The provided code contains pre-trained weight.

\begin{table*}
\begin{center}
\label{table:headings}
\renewcommand{\tabcolsep}{4pt}
\makeatletter
\def\hlinewd#1{%
\noalign{\ifnum0=‘}\fi\hrule \@height #1 \futurelet
\reserved@a\@xhline}
\newcommand{\hthickline}{\hlinewd{1pt}}
\newcommand{\hthinline}{\hlinewd{.2pt}}
\makeatother
\newcolumntype{Z}{>{\centering\arraybackslash}X}
{\small
\begin{tabularx}{\linewidth}{Z||Z|Z|c|c|c}
\hthickline
Network &Layer (BN, activation function) &Output size &Kernel &Stride &Pad\\
\hline\noalign{\smallskip}
\hline
\multirow{1}{*}{Encoder}  &Resnet-18 &8 x 8 x 512 & - & - & - \\
\hline

\multirow{12}{*}{Decoder}  
&Conv 1 (BN, ReLU) &8 x 8 x 512 &3 x 3 &1 &1\\
&ConvTr 1 (BN, ReLU) &16 x 16 x 512 &4 x 4 &2 &1\\
&Conv 2 (BN, ReLU) &16 x 16 x 256 &3 x 3 &1 &1\\
&ConvTr 2 (BN, ReLU) &32 x 32 x 256 &4 x 4 &2 &1\\
&Conv 3 (BN, ReLU) &32 x 32 x 128 &3 x 3 &1 &1\\
&ConvTr 3 (BN, ReLU) &64 x 64 x 128 &4 x 4 &2 &1\\
&Conv 4 (BN, ReLU) &64 x 64 x 128 &3 x 3 &1 &1\\
&ConvTr 4 (BN, ReLU) &128 x 128 x 128 &4 x 4 &2 &1\\
&Conv 5 (BN, ReLU) &128 x 128 x 128 &3 x 3 &1 &1\\
&ConvTr 5 (BN, ReLU) &256 x 256 x 128 &4 x 4 &2 &1\\
&Conv 6 (BN, ReLU) &256 x 256 x 128 &3 x 3 &1 &1\\
&Conv 7 (-, Sigmoid) &256 x 256 x 3 &3 x 3 &1 &1\\

\hline
\multirow{8}{*}{Discriminator}  
&Conv 1 (-, LeakyReLU) &128 x 128 x 64 &4 x 4 &2 &1\\
&Conv 2 (BN, LeakyReLU) &64 x 64 x 128 &4 x 4 &2 &1\\
&Conv 3 (BN, LeakyReLU) &32 x 32 x 256 &4 x 4 &2 &1\\
&Conv 4 (BN, LeakyReLU) &16 x 16 x 512 &4 x 4 &2 &1\\
&Conv 5 (BN, LeakyReLU) &8 x 8 x 512 &4 x 4 &2 &1\\
&Conv 6 (BN, LeakyReLU) &4 x 4 x 512 &4 x 4 &2 &1\\
&Conv 7 (BN, LeakyReLU) &2 x 2 x 128 &4 x 4 &2 &1\\
&Conv 8 (-, Sigmoid) &1 x 1 x 1 &4 x 4 &2 &1\\
\hline
\end{tabularx}}
\end{center}
\caption{Architectural details of the proposed method. ConvTr denotes a transposed convolution layer and Conv denotes a convolution layer.}
\end{table*}

\section{Analysis of Threshold Sensitivity}

In this section, we show the IoU results according to threshold changes for each category in the MVTec AD dataset. As shown in Figs. 10, 11, and 12, compared to SPADE and Patch SVDD, which are comparative methods, the performance difference of the proposed AnoSeg is not large according to the change in the threshold.

\begin{figure}[b]
\begin{center}
\includegraphics[width=1.0\linewidth]{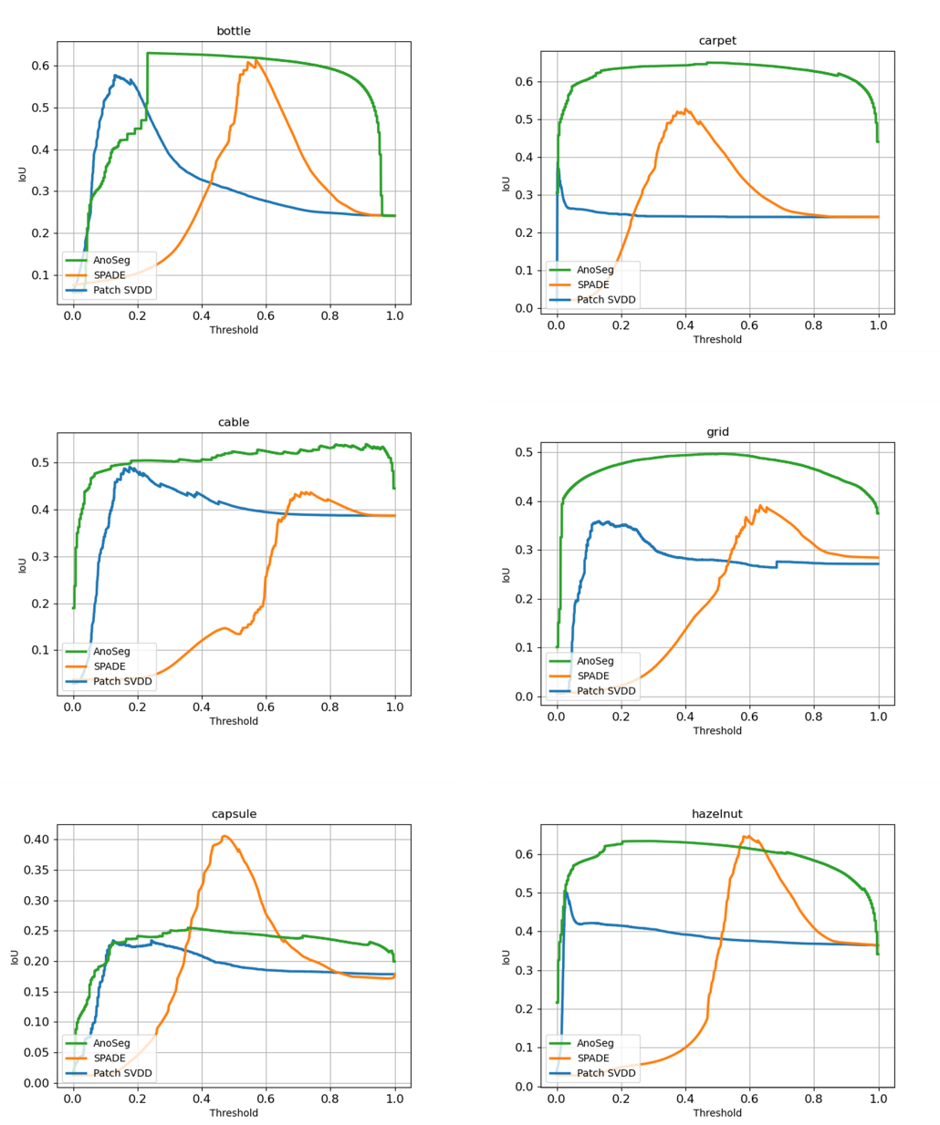} 
\end{center}

  \caption{IoU results for each category in the MVTec AD dataset according to the threshold change. (Green: AnoSeg, Orange: SPADE, Blue: Patch SVDD)}
\label{fig9}
\end{figure}

\begin{figure}[t]
\begin{center}
\includegraphics[width=1.0\linewidth]{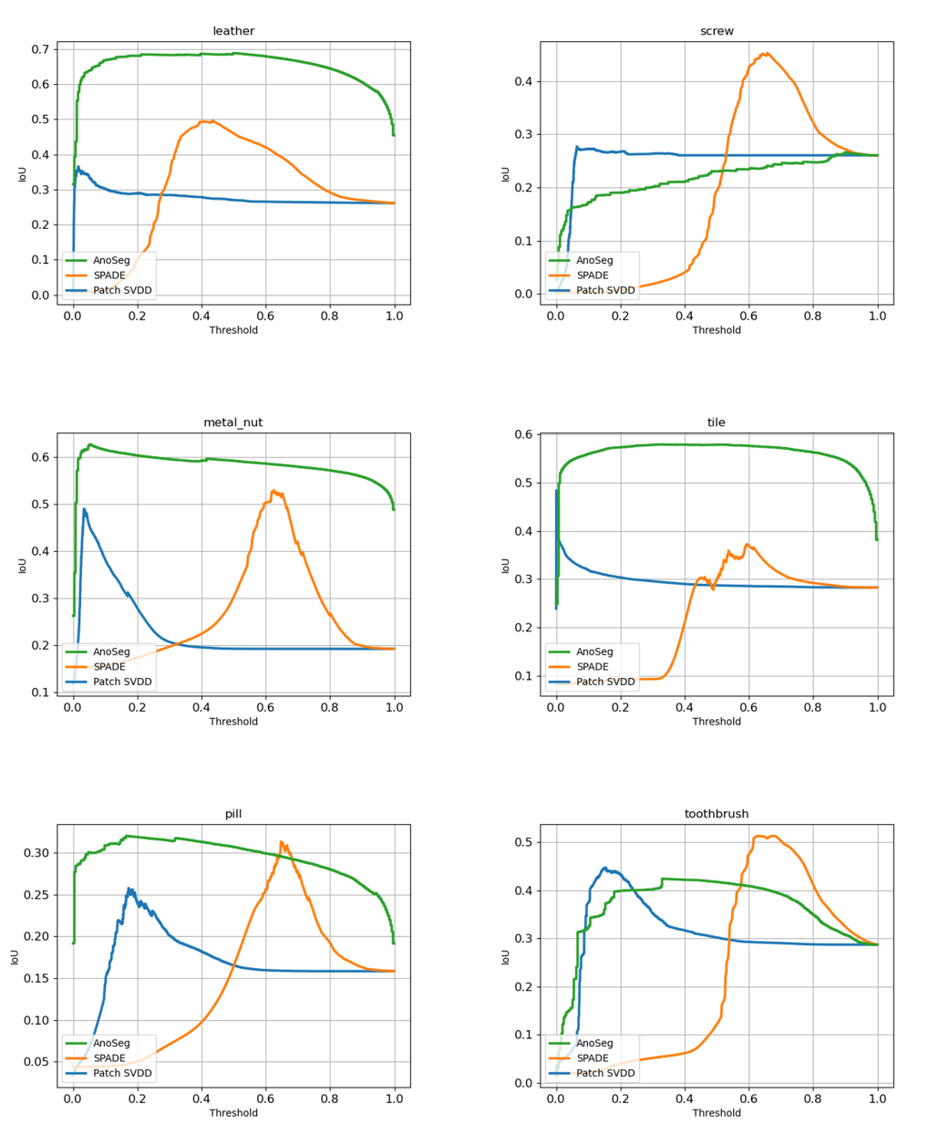} 
\end{center}

  \caption{IoU results for each category in the MVTec AD dataset according to the threshold change. (Green: AnoSeg, Orange: SPADE, Blue: Patch SVDD)}
\label{fig10}
\end{figure}

\begin{figure}[t]
\begin{center}
\includegraphics[width=1.0\linewidth]{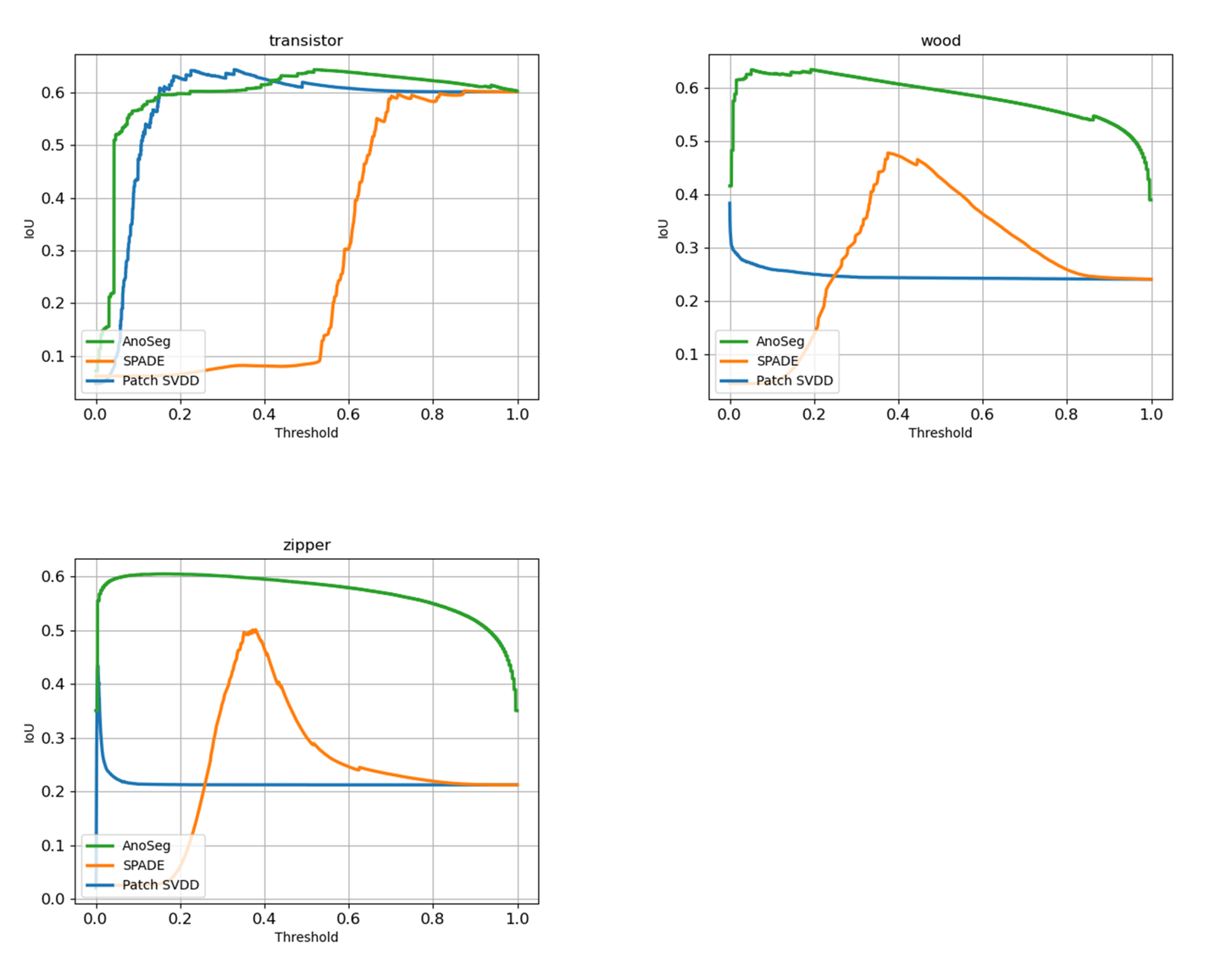} 
\end{center}

  \caption{IoU results for each category in the MVTec AD dataset according to the threshold change. (Green: AnoSeg, Orange: SPADE, Blue: Patch SVDD)}
\label{fig11}
\end{figure}

\section{Qualitative results on the MVTec AD dataset}

We provided additional qualitative results of our method on the MVTec AD dataset in Figs. 13, 14, 15, 16, and 17. For each class, an Input image, a proposed anomaly map, and a GT mask are provided. The proposed AnoSeg had the highest performance even for anomaly regions with various sizes.

\begin{figure}[t]
\begin{center}
\includegraphics[width=1.0\linewidth]{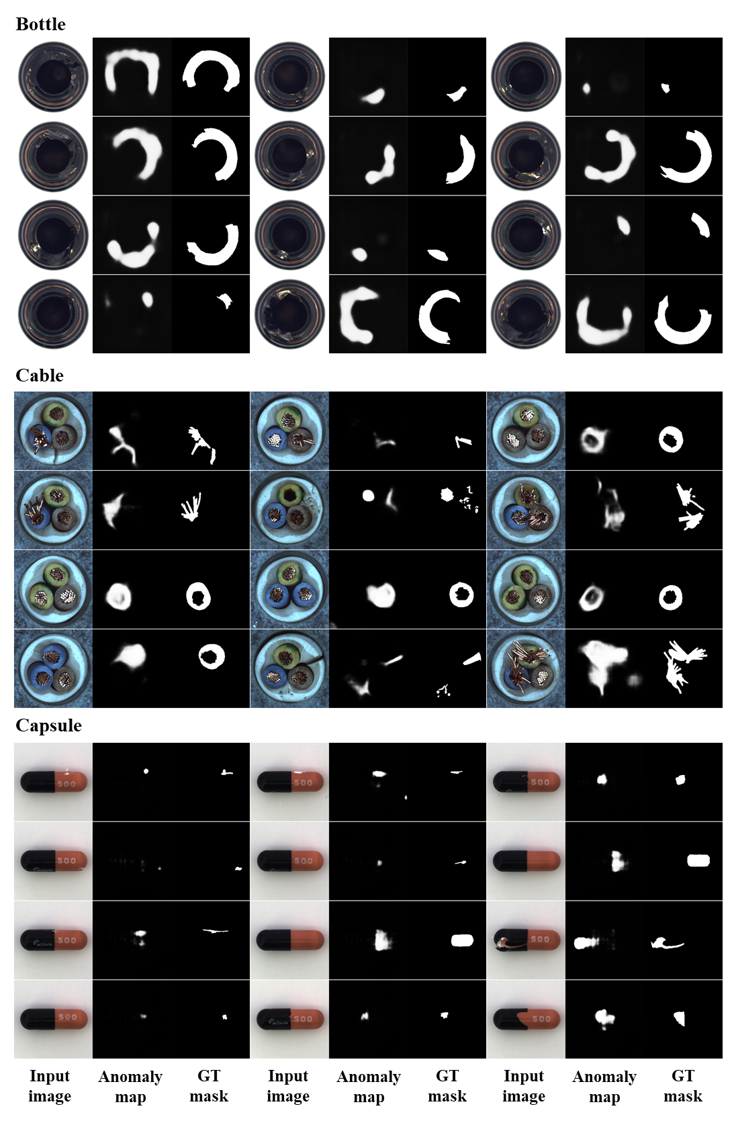} 
\end{center}

  \caption{Defect segmentation on MVTec AD dataset. For each sample image, there are an input image, the proposed anomaly map, and its GT mask from left to right.}
\label{fig12}
\end{figure}

\begin{figure}[t]
\begin{center}
\includegraphics[width=1.0\linewidth]{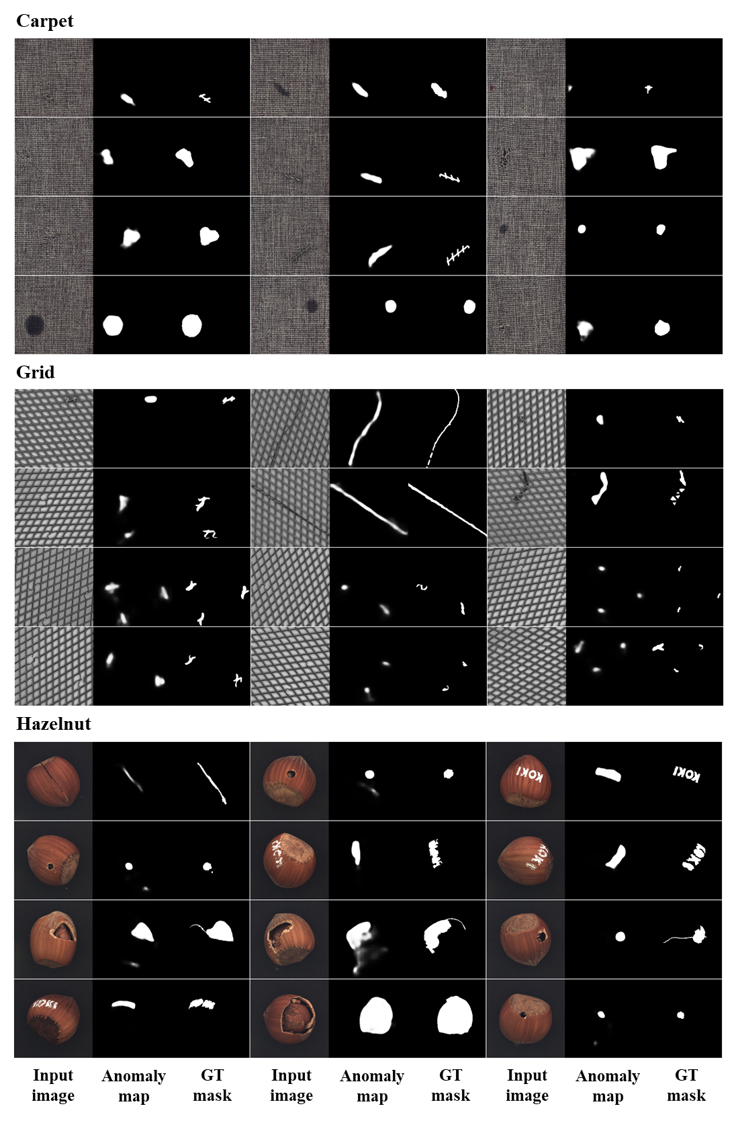} 
\end{center}

  \caption{Defect segmentation on MVTec AD dataset. For each sample image, there are an input image, the proposed anomaly map, and its GT mask from left to right.}
\label{fig13}
\end{figure}

\begin{figure}[t]
\begin{center}
\includegraphics[width=1.0\linewidth]{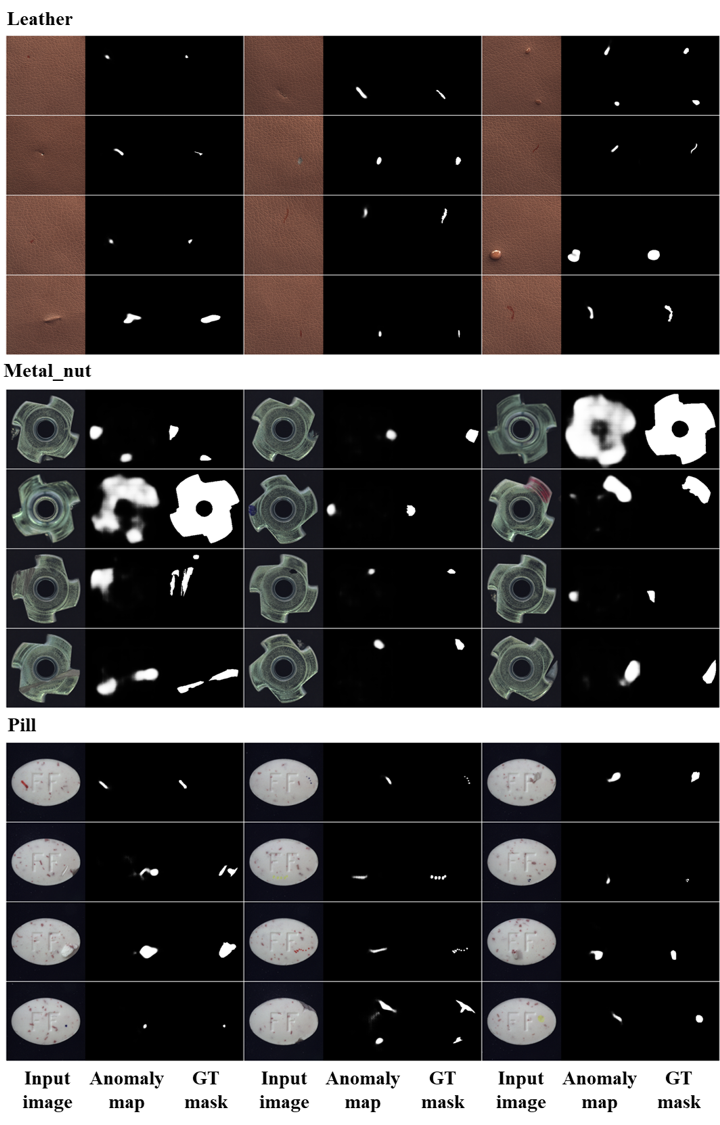} 
\end{center}

  \caption{Defect segmentation on MVTec AD dataset. For each sample image, there are an input image, the proposed anomaly map, and its GT mask from left to right.}
\label{fig14}
\end{figure}

\begin{figure}[t]
\begin{center}
\includegraphics[width=1.0\linewidth]{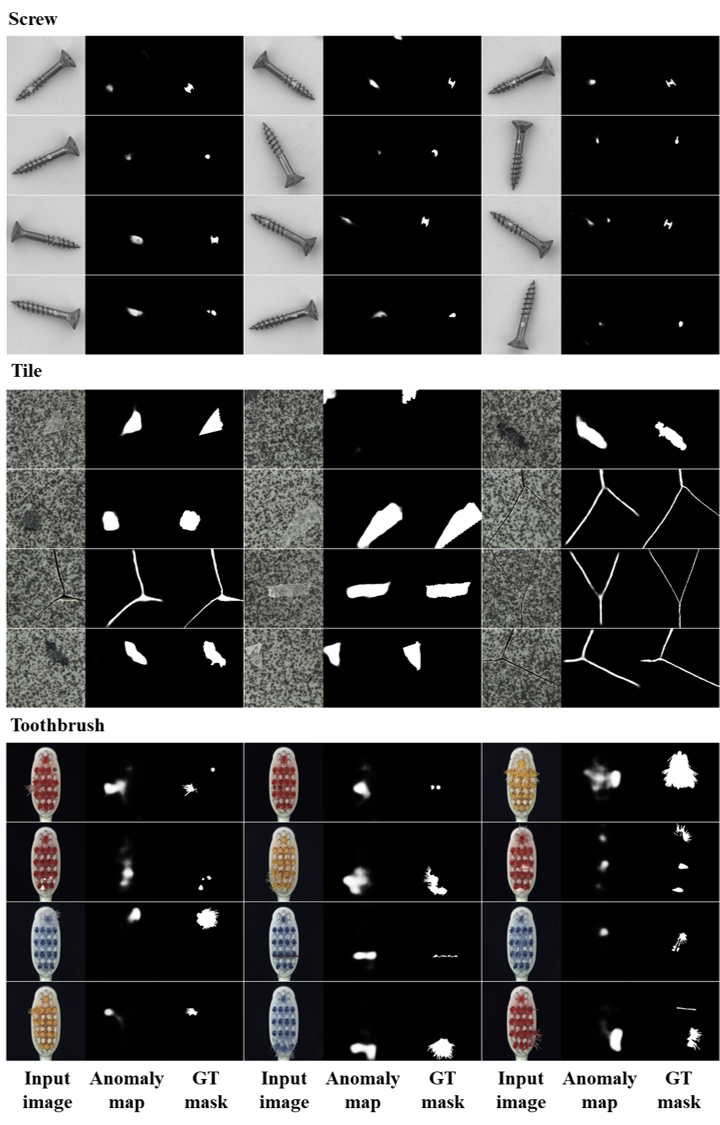} 
\end{center}
  \caption{Defect segmentation on MVTec AD dataset. For each sample image, there are an input image, the proposed anomaly map, and its GT mask from left to right.}
\label{fig15}
\end{figure}

\begin{figure}[t]
\begin{center}
\includegraphics[width=1.0\linewidth]{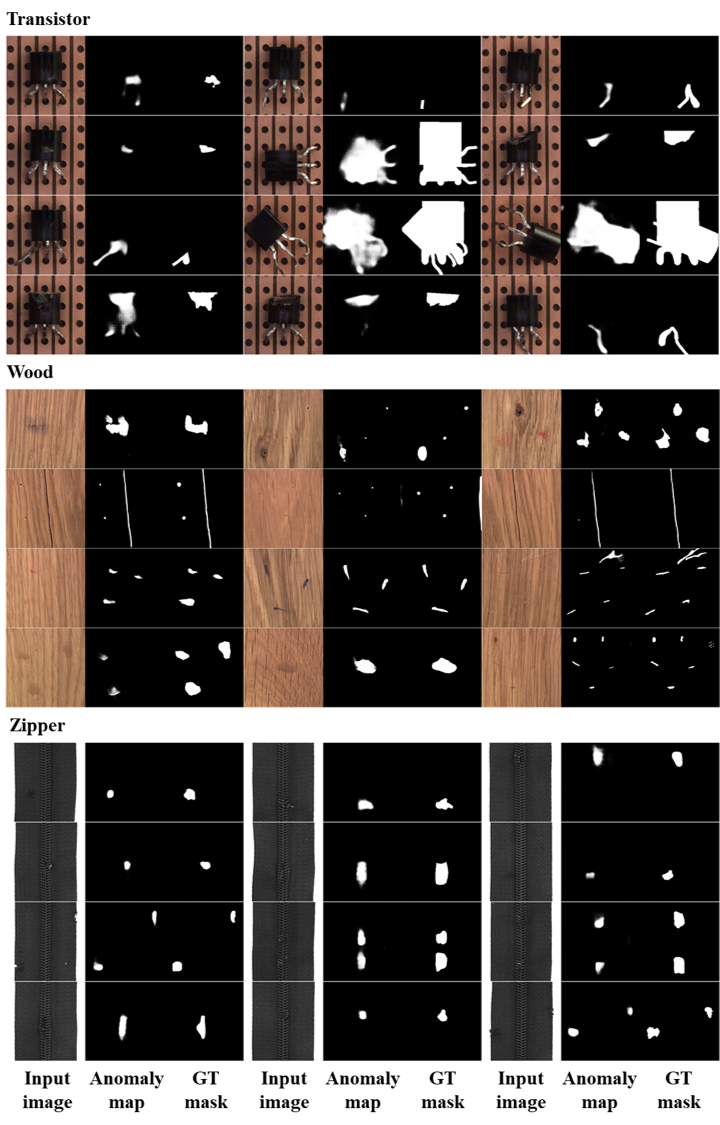} 
\end{center}

  \caption{Defect segmentation on MVTec AD dataset. For each sample image, there are an input image, the proposed anomaly map, and its GT mask from left to right.}
\label{fig16}
\end{figure}

\end{document}